\documentclass{article}

\pdfoutput=1

\usepackage{natbib}

\usepackage{graphicx}
\usepackage{mathptmx}      
\usepackage{commath}
\usepackage{tikz}
\usetikzlibrary{calc,intersections,through,backgrounds}
\usetikzlibrary{shapes,arrows}
\usepackage{amsmath,amssymb,stmaryrd,textcomp}
\usepackage{booktabs} 
\usepackage{verbatim}

\usepackage[normalem]{ulem}

\tikzstyle{cm dotted}=[dash pattern=on \pgflinewidth off .25mm ]
\tikzstyle{decision} = [diamond, draw, fill=blue!10, 
text width=4.5em, text badly centered, node distance=3cm, inner sep=0pt]
\tikzstyle{block} = [rectangle, draw, fill=blue!10, 
text width=5em, text centered, rounded corners, minimum height=4em]
\tikzstyle{bigblock} = [rectangle, draw, text width=5em, text centered, rounded corners, minimum height=2cm,minimum width=5cm]
\tikzstyle{contblock} = [rectangle, draw, dashed, text width=5em, text centered, rounded corners, minimum height=3.75cm,minimum width=2.75cm]
\tikzstyle{cablock} = [rectangle, draw, dashed, text width=5em, text centered, rounded corners, minimum height=8.75cm,minimum width=6cm]
\tikzstyle{arrow} = [draw, -latex']
\tikzstyle{line} = [draw,-]
\tikzstyle{cloud} = [draw, ellipse,fill=red!20, node distance=3cm,
minimum height=2em]

\begin{document}

\title{A hybrid cellular automaton model of cartilage regeneration capturing the interactions between cellular dynamics and scaffold porosity}
\author{ {\sc Simone Cassani and Sarah D. Olson}\\[2pt]
Department of Mathematical Sciences, Worcester Polytechnic Institute, \\
100 Institute Rd, Worcester MA 01609, USA.\\[6pt]
}
\pagestyle{headings}
\markboth{S. CASSANI}{\rm Hybrid model of cartilage regeneration}
\maketitle

\begin{abstract}
To accelerate the development of strategies for cartilage tissue engineering, models are necessary to investigate the interactions between cellular dynamics and scaffold porosity. In experiments, cells are seeded in a porous scaffold or hydrogel where over the time course of a month, the scaffold slowly degrades while cells divide and synthesize extracellular matrix constituents. We use an off-lattice cellular automaton framework to model the individual behavior of cells within the scaffold. The movement of cells and the ability to reproduce is determined by the nutrient profile in the construct and/or the local porosity, defined as the volume fraction of fluid in the surrounding region. A phenomenological approach is used to capture a continuous profile for the degrading scaffold and accumulating matrix, which will then change the local porosity throughout the construct. We parameterize the model by first matching total cell counts in the construct to chondrocytes seeded in a polyglycolic acid scaffold \citep{Freed94}.  We investigate the total cell count and location of different cell populations within the construct for different initial scaffold porosities. Similar to experiments, we observe cell counts that level off around day 15 with higher cell counts in scaffolds of higher initial solid volume fraction (lower porosity).  Cell clustering is observed and characterized in regions at the edge of the construct that are close to the nutrient-rich medium in the fluid bath. Model results show that a bias in motion due to a sensitivity to porosity allows cells to move in a smarter or more optimal arrangement. We investigate the spatiotemporal distribution of cells as the cell reproduction rate, cell movement distance, and sensitivity to porosity is varied.  We observe non monotonic changes in total cell counts within different regions of the construct due to the interplay between porosity and cellular movement. We also analyze the emergent average cell speed for different initial scaffold porosities, observing higher average cell  speed over the course of 30 days for the lowest initial scaffold porosity. 
This model provides a framework to further investigate how changes in biological parameters such as cell division and movement can change the cellular count and distribution in scaffolds of different initial porosity. \medskip 
\end{abstract}

\section{Introduction}\label{intro}
Articular cartilage (AC) is an avascular and aneural connective tissue that covers articular joints such as the shoulder and knees to provide a viscoelastic surface that distributes and absorbs mechanical loads. The complex structure of the AC consists of cells called chondrocytes and a dense extracellular matrix (ECM) including fluid, a collagen network, and other proteins~\citep{Mak86,Mow92,Wong03}. Chondrocytes are responsible for the production of the building blocks of the ECM~\citep{Archer03,Akkiraju15,Muir95}, while nutrients and oxygen are provided via diffusion through the ECM. Erosion and damage of AC can be caused by injuries, mechanical wear, and pathologies such as osteoarthritis. Cartilage degeneration from osteoarthritis is often painful and can affect people of all ages (including the majority of the 65+ population). The absence of vasculature and the low density and metabolism of chondrocytes make cartilage a tissue with low capability for repair. Common invasive strategies to increase mobility include joint replacements, which include a long recovery time ~\citep{Goldring07,Kock12,Mollenhauer08}.  Hence, tissue engineering represents a promising path towards the treatment of damaged cartilage that could be non-invasive  with immediate relief.

Tissue engineering is a multidisciplinary field in the area of tissue repair and regeneration. Cells can be seeded on a three dimensional porous scaffold, where nutrients and oxygen are provided, to produce new tissue that can be successfully used to repair damaged areas and aid in tissue regeneration. Experiments producing tissue engineered constructs are costly, time intensive and must produce viable and functional tissue to be considered successful. The tissue produced in experiments requires the correct mechanical and biological properties in order to survive when implanted at the level of the trauma and to withstand the stresses of the affected region. Thus, nutrients must be able to diffuse throughout the construct and cells must be able to synthesize the proteins that will bind to form functional ECM, providing mechanical support as the scaffold slowly degrades. Scientists have identified that the cell source, type of mechanical stimulation, interactions between the scaffold and cells, delivery of growth factors, and scaffold properties are all important parameters for tissue engineering experiments~\citep{Freyria04,Freyria12,Keeney11,Kock12,Kretlow08,Reddi11}. 

It is therefore imperative that scaffolds are able to mimic the properties of the ECM, providing an environment that allows cells to divide and move. In addition, it is necessary that the scaffold also degrades at the proper timescale in order to obtain a valid outcome from experiments~\citep{Fernandes09}. 
As a result, several recent studies have focused on the effect of scaffold pore size and micro architecture in different  tissue engineering experiments~\citep{Erickson09,Matsiko15,Reddi11}.
In particular, studies on tissue engineered articular cartilage have investigated the effect of variations in experimental parameters, such as: scaffold porosity~\citep{Erickson09,Freyria04,Matsiko15}, scaffold thickness~\citep{Freed94}, initial cellular seeding concentration~\citep{Cigan16,Freed94,Vunjak98}, mixed and static cultures~\citep{Freyria04,Vunjak98}, and the presence of nutrient channels~\citep{Cigan16}, on the biological and mechanical properties of the construct produced {\it in vitro}. 

Mathematical modeling can be used alongside experiments to help in the interpretation of laboratory results, to gain a deeper understanding of the mechanisms driving the dynamics of cartilage growth, and to quantify the effect of variations in the experimental parameters. Tissue growth has been studied from several mathematical modeling frameworks in the context of tissue engineering as well as cancer tumor growth \citep{AlHusari14,Galle07,Sengers07,Zhang09}.  \cite{Cheng06} built a discrete on-lattice three-dimensional cellular automaton model describing cell dynamics assuming constant concentration of nutrient and growth factors. The model was later coupled to a time dependent partial differential equation for the diffusion and consumption of nutrient and to an equation regulating cell division time and migration speed in a hybrid model by \cite{Cheng09}. \cite{Jeon2010} developed a hybrid off-lattice two-dimensional model for tumor growth adopting a continuum approach for enzymes, nutrient and ECM concentration and a discrete approach for individual cellular behavior. We note that in \cite{Byrne09}, a comparison between a lattice-free, rule-based cellular automaton and a continuum model showed good agreement, showing that both modeling frameworks can be utilized for tissue growth where there is a growing cell population. 

Mathematical models specific to cartilage tissue engineering have been focusing on different elements of the biological phenomenon of cartilage growth, using a variety of modeling techniques. The two-dimensional hybrid model by \cite{Chung10} combines a partial differential equation for the concentration of nutrients in the scaffold with an on-lattice cellular automaton model for cell random movement, nutrient regulated cell reproduction and cell-to-cell interaction. To our knowledge, this was the first model to account for individual cell dynamics in the context of cartilage tissue engineering, although this model did not account for the evolving ECM and scaffold volume fractions in the construct. Other groups have developed continuous mixture model approaches to capture the evolving solid and fluid volume fractions \citep{Chung06,Haider11,Lemon06,ODea13}, where cells were either assumed constant or accounted for via a cellular volume fraction. There have also been a series of continuous models that have focused on accounting for spatiotemporal distribution of cells as well as the biosynthesis of ECM proteins and diffusion of nutrients \citep{Dimicco03,Nikolaev10,Obradovic00,Sengers04,Sengers05a,Sengers05b}. 

In this work, we present the first hybrid model to account for the effect of porosity in a cartilage tissue engineered construct. The model couples a cellular automaton description of the chondrocytes (cells) to a continuum phenomenological approach for ECM accumulation, scaffold degradation, and nutrient concentration. Chondrocytes are modeled as exhibiting biased random motion depending on local nutrient concentration (chemotaxis) and porosity, and their division is regulated by nutrients and porosity. The model is used to investigate how changes in biological parameters such as distance a cell moves, sensitivity to porosity, and cell maturity affect constructs with a different initial porosity, and to aid in the interpretations of the results of laboratory experiments on tissue engineered articular cartilage. We are able to match total cell counts of experiments by \cite{Freed94} and show that cell's sensitivity to porosity can lead to different cell counts and distribution of cells within the construct. This work provides a new framework to couple porosity with individual cellular dynamics for the growth of tissue in cell-seeded scaffolds.

\section{Modeling Framework}
\label{sec:methods}
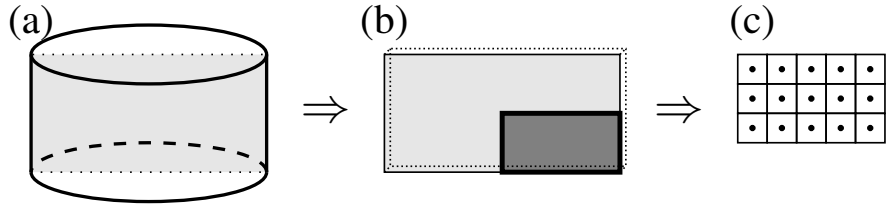
\begin{figure}[th!]
	\centering
	\resizebox{\textwidth}{!}{ 
		\begin{tikzpicture}
		\node [above] (a) at (-1,0.5) {(a)}; 
		\draw [dotted,fill=gray!20] (-1,-0.5) rectangle (1,0.5);
		\draw [thick] (-1,-0.5) -- (-1,0.5);
		\draw [thick] (1,-0.5) -- (1,0.5);
		\draw [thick] (0,0.5) ellipse (1cm and 0.25cm);
		\draw [thick] (-1,-0.5) arc (180:360:1cm and 0.25cm);
		\draw [thick,dashed] (1,-0.5) arc (0:180:1cm and 0.25cm);
		\node [xshift=0.5cm] (AA) at (1,0) {\large$\Rightarrow$};
		\begin{scope}[xshift=3cm]
		\node [above] (b) at (-1,0.5) {(b)};
		\draw [fill=gray!20] (-1,-0.5) rectangle (1,0.5); 
		\draw [very thick,fill=gray] (0,0) rectangle (1,-0.5);
		\draw [thin,cm dotted] (-1,0.5) -- (-0.95,0.55) -- (1.05,0.55)-- (1.05,-0.45) -- (1,-0.5);
		\draw [thin,cm dotted] (-0.95,0.55) -- (-0.95,-0.45) -- (1.05,-0.45);
		\draw [thin,cm dotted] (1,0.5) -- (1.05,0.55);
		\draw [thin,cm dotted] (-1,-0.5) -- (-0.95,-0.45);
		\end{scope}
		\node [xshift=3.5cm] (AA) at (1,0) {\large$\Rightarrow$};
		\begin{scope}[xshift=5.25cm]
		\node [above] (c) at (-0.125,0.5) {(c)};
		\foreach \y in {-0.125cm,0.125cm,0.375cm}{
			\foreach \x in {-0.125cm,0.125cm,...,0.875cm}{
				\begin{scope}[xshift=\x,yshift=\y]
				\draw [thin] (-0.125,-0.125) rectangle (0.125,0.125);
				\draw [fill=black] (0,0) circle (0.02cm);
				\end{scope}
		}}
		%
		\end{scope}
		\end{tikzpicture} }
	\caption{Schematic of the computational domain. (a): Cylindrical scaffold with the middle slice highlighted in light gray. (b): The computational domain $\Omega$ is one-quarter of this thin slice, corresponding to the dark gray rectangle in the lower right corner. (c): Zoomed in schematic of the domain $\Omega$ where the node values correspond to averages on the particular volume element with depth $z=2c_d$ for chondrocyte diameter $c_d$.}
	\label{fig:sc}
\end{figure}

This model describes the phenomena of growth and development of tissue engineered articular cartilage where initially, chondrocytes are seeded in a cylindrically shaped scaffold or gel. As depicted in Fig.~\ref{fig:sc}(a)-(b), we focus on a thin slice of this construct and only model the lower right corner since we assume the entire construct is immersed in a well-mixed, nutrient-rich medium  where all surfaces of the cylinder are fully exposed to allow for nutrient diffusion into the construct. We note that the average diameter $c_d$ of a chondrocyte is  $c_d=0.001$ cm~\citep{Sanchez10}  whereas average values for the height $h$ and diameter $d$ of cylindrical scaffolds in experiments are $h=0.2\ll c_d$ and $d=1\ll c_d$ cm \citep{Erickson09,Freed94}, respectively.  The thin slice is assumed to have thickness $z=$2$c_d$ and we use dimensional analysis to arrive at a 2-dimensional domain $\Omega$, where values of continuous variables correspond to cross sectional averages of the regions depicted in Figure~\ref{fig:sc}(c). Since the thickness is 2$c_d$, we assume that cells can partially overlap and that there is room for scaffold and/or ECM at each point in $\Omega$.

Experimental studies have shown that movement of cells and diffusion of nutrients will be heavily dependent on the porosity (fluid volume fraction) that is not obstructed by the proteins or fibers that make up the scaffold or extracellular matrix \citep{Erickson09,Freyria04,Masaro99,Matsiko15}. Thus, we will idealize the bio-construct as a continuum mixture where we will account for the evolving solid volume fractions and hence, the porosity. 
The model employs a hybrid approach that combines a discrete {\it cellular automaton} (CA) model for the cells with continuous descriptions of the nutrient concentration ($c$), the porosity ($p$) of the construct, the extracellular matrix solid volume fraction ($\Phi_{ECM}$), the scaffold solid volume fraction ($\Phi_{SC}$), and cellular solid volume fraction ($\Phi_{C}$). We assume a saturated mixture for the entire experiment, which corresponds to the assumption that at each point in the construct, the porosity is $p=1-\Phi$ for total solid volume fraction $\Phi=$(occupied volume)/(total volume). In this model, we assume that nutrients are dissolved in the fluid phase and do not contribute to the solid volume fraction.

\begin{figure}[th!]
	\centering
	\includegraphics[width=0.95\textwidth]{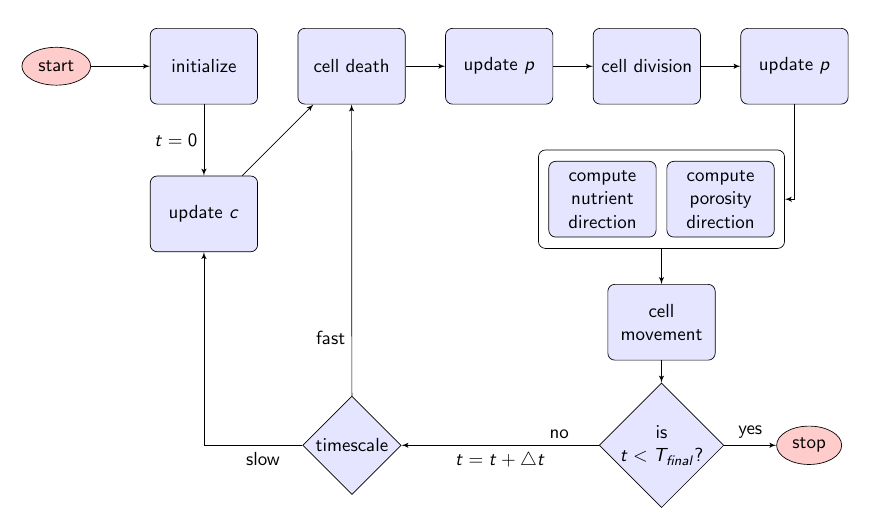}
	\caption{Algorithm for the hybrid model where cells are represented using a Cellular Automaton approach to account for growth and movement of cells. The nutrient concentration $c$ and porosity $p$ are represented in a continuous framework. }
	\label{fig:flux-big}
\end{figure}

The diagram in Fig.~\ref{fig:flux-big} describes the algorithm that is used to simulate the evolution of the bio-construct in time. At the system initialization, the cells are seeded on the domain; then the nutrient part of the {\it continuous block} is solved, updating the value of nutrient concentration inside the computational domain. Next, the first part of the {\it CA block} is solved and cells in the dying state are removed from the domain, while the dying cells for the next iteration are randomly selected. Then, the second part of the continuous block is solved to perform the update of the construct porosity due to changes induced by cell death and cell movement (for $t>0$). Next, the second part of the CA block is solved to perform cell division and the second part of the continuous block is solved again to update the construct porosity due to changes induced by cell division. Finally, the second part of the CA block is solved to perform cell movement. This algorithm is repeated until the simulation time $t$ reaches $T_{final}$. We note that the nutrient concentration $c$ is updated only at $t=0$, 7, 10, and 15 days (slow timescale) as detailed in Appendix B. In the following sections, the different components of the model will be discussed in detail.

\subsection{Continuous model}

The continuous components of the model describe the evolution of the nutrient concentration $c$ and porosity $p$. These quantities are calculated on a computational grid with a step size $\triangle x=\triangle y=c_d/12$, i.e. the grid is considerably smaller then the cellular diameter to guarantee an accurate solution. We assume that the nutrient profile will only change after a significant change in the distribution of cells within the construct (both in terms of the total number of cells and their respective locations in the domain). Thus, the two continuous quantities considered are computed on different timescales: $c$ is considered independent from the cellular dynamics, with a slow time evolution compared to the cells while $p$ directly depends on the cellular dynamics such as movement and growth, and is therefore computed on the fast ({\it cellular}) time scale.  

\subsubsection{Nutrient concentration} 
For the domain highlighted in Fig.~\ref{fig:sc}(b), nutrients (e.g. glucose) will diffuse in from the bottom and right since they are in contact with the nutrient-rich medium. We will utilize a qualitative approach to specify $c$, the concentration of nutrient solute per mixture volume (equivalent to $p\hat{c}$ for porosity $p$ and concentration of nutrient solute per solute volume $\hat{c}$). The spatial profile of $c$ is chosen in agreement with the results of \cite{Chung10} and \cite{Bandeiras15}. As the simulation time evolves, the nutrient concentration in the areas far from the physical boundary of the bio-construct decrease significantly, as shown in Fig.~\ref{fig:nutrient}. This represents cell agglomerates that form close to the boundary (bottom and right of $\Omega$ in Fig.~\ref{fig:sc}), where there is increased cellular utilization of nutrients and a decrease in nutrient diffusivity due to a reduction in porosity. Since we assume the timescale characteristic of this process is slow compared with the timescale governing the cellular processes, $c$ is updated on the {\it slow} timescale corresponding to new profiles on day $0$, $7$, $10$ and $15$. Additional details about the specific profiles used in this work are reported in Appendix B. We note that we utilize a non-dimensional nutrient concentration with $c\in$[0,1] since we want all variables that bias the cellular components to be on the same scale.

\begin{table}[t!]
	\begin{center}
		\begin{tabular}{llcl}
			\toprule
			\bf Parameter (units) & \bf Description & \bf Sim. Value & \bf Source \\
			\midrule
			$c_d$ (cm) & Diameter of chondrocyte  & $0.001$ & \cite{Sanchez10} \\
			$\Omega$ (cm) & Domain & $0.5\times 0.1$  & \cite{Erickson09} \\
			& & & \cite{Freed94}\\
			$z$ (cm) & Thickness of thin slice & 2$c_d$ & see Appendix A\\
			$\triangle x$, $\triangle y$ & Grid spacing on domain $\Omega$ & $c_d/12$ & this work \\
			$\triangle t$ (hrs)& Time step & 0.5 & this work\\
			$\widehat{\Phi}$ & Cellular vol. frac. at a node & 0.3 & see Appendix A\\
			$\Phi_{SC}^0$ & Initial SC vol. frac. & [0.01, 0.05] & \cite{Erickson09} \\
			& & & \cite{Freed94}\\
			$k_{SC}$ (1/days) & Degradation rate of SC & 0.038 & \cite{Wilson02}\\
			$\tau$ (days) & Delay for SC degradation & 14 & \cite{Freed94}\\
			$\Phi_{ECM}^{SS}$ & ECM steady state vol. frac.  & 0.15 &\cite{Podrazky66}\\
			$k_{ECM}$ (1/days) & ECM growth rate  & 0.05 & \cite{Wilson02}\\
			$t_{div}$ (days) & Minimum cell age to div. & 2.5 & \cite{Bandeiras15} \\
			& & &\cite{Freed94} \\
			& & & \cite{Chung10}\\
			$t_{idle}$ (hrs) & Cell idle time after collision & 4 & \cite{Chung10}\\
			$d_{cell}$ (cm) & Dist. a cell moves in $\triangle t$ & 0.0005 & \cite{Morales07}  \\
			$p_{sens}$ & Min. porosity for div. or mov. & 0.6 & see Appendix A\\
			$a_{d,mov}$ (cm) & Diam. of annulus for cell mov. & 3$c_d$ & see Appendix A \\
			$d_{div}$ (cm) & Min. dist. for div. cells & 3.5$c_d$ &  this work\\
			$c_{div}$ & Min. mean $c$ for div. & 0.7 & this work \\
			$a_{d,div}$ (cm) & Diam. of annulus for cell div. & 3.5$c_d$ & see Appendix A\\
			$p_{death}$& Probability of dying per $\triangle t$ & 0.0006 &  \cite{Bandeiras15} \\
			\bottomrule
		\end{tabular}
		\caption{Parameters used in the model to describe time scales and rates for cell growth and movement, as well as extracellular matrix accumulation and scaffold degradation. Notation -  ECM=extracellular matrix, SC=scaffold, Sim.=simulation, dist.=distance, mov.=movement, div.=dividing, min.=minimum, vol. frac.=volume fraction, diam.=diameter, $c$=non dimensional nutrient concentration.}
		\label{tab:par}
	\end{center}
\end{table}

\subsubsection{Porosity} 
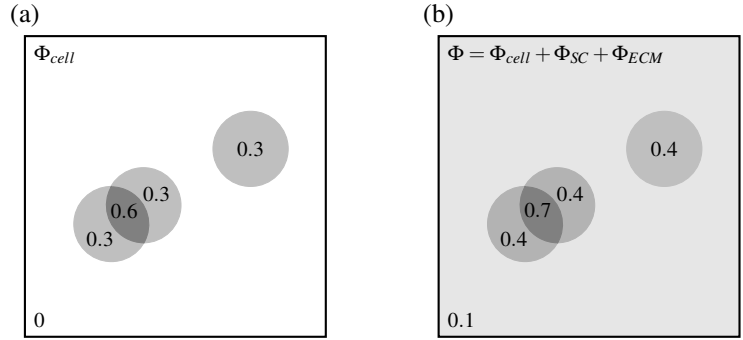
\begin{figure}[ht!]
	\centering
	\begin{tikzpicture}
	\node[above] (A) at (-2,2) {(a)};
	\draw [thick] (-2,-2) rectangle (2,2);
	\draw [draw=gray!50,fill=gray!50] (-0.85,-0.5) circle (0.5cm);
	\draw [draw=gray!50,fill=gray!50] (-0.425,-0.25) circle (0.5cm);
	\draw [draw=gray,fill=gray] (-0.42,-0.755) arc (-30:90:0.5cm) --
	(-0.85,0) arc (150:270:0.5cm);
	\draw [draw=gray!50,fill=gray!50] (1,0.5) circle (0.5cm);
	\node[below right] (A) at (-2,2) {\footnotesize $\Phi_{cell}$};
	\node (A) at (-0.25,-0.1) {\footnotesize $0.3$};
	\node (A) at (-0.68,-0.32) {\footnotesize $0.6$};
	\node (A) at (-1,-0.7) {\footnotesize $0.3$};
	\node (A) at (1,0.5) {\footnotesize $0.3$};
	\node[above right] (A) at (-2,-2) {\footnotesize $0$};
	\begin{scope}[xshift=5.5cm]
	\node[above] (A) at (-2,2) {(b)};
	\draw [thick,fill=gray!20] (-2,-2) rectangle (2,2);
	\draw [draw=gray!60,fill=gray!60] (-0.85,-0.5) circle (0.5cm);
	\draw [draw=gray!60,fill=gray!60] (-0.425,-0.25) circle (0.5cm);
	\draw [draw=gray,fill=gray] (-0.42,-0.755) arc (-30:90:0.5cm) --
	(-0.85,0) arc (150:270:0.5cm);
	\node[below right] (A) at (-2,2) {\footnotesize $\Phi=\Phi_{cell}+\Phi_{SC}+\Phi_{ECM}$};
	\draw [draw=gray!50,fill=gray!50] (1,0.5) circle (0.5cm);
	\node[above right] (A) at (-2,-2) {\footnotesize $0.1$};
	\node (A) at (-0.25,-0.1) {\footnotesize $0.4$};
	\node (A) at (-0.68,-0.32) {\footnotesize $0.7$};
	\node (A) at (-1,-0.7) {\footnotesize $0.4$};
	\node (A) at (1,0.5) {\footnotesize $0.4$};
	\end{scope}
	\end{tikzpicture}
	\caption{Examples of solid volume fractions for different components in the model. (a): The cellular solid volume fraction $\Phi_{cell}$ is shown for the case where we assume each cell contributes $\widehat{\Phi}=0.3$ to the solid volume fraction. (b): Total solid volume fraction ${\Phi=\Phi_{cell}+\Phi_{SC}+\Phi_{ECM}}$ for the same cellular locations with $\Phi_{SC}+\Phi_{ECM}=0.1$. The domain $\Omega$ assumes a thickness that allows for partial overlap of cells and for scaffold and/or extracellular matrix around the cells.}
	\label{fig:por}
\end{figure}

The porosity $p$ in the bio-construct is defined as $1-\Phi$ for solid volume fraction $\Phi$. At each node of the computational grid, $\Phi$ is computed as the sum of all the solid volume fractions: the cellular volume fraction $\Phi_{cell}$, the scaffold volume fraction $\Phi_{SC}$ and the extracellular matrix volume fraction $\Phi_{ECM}$, (ie $\Phi=\Phi_{cell}+\Phi_{SC}+\Phi_{ECM}$). This is illustrated in Fig.~\ref{fig:por}(b) for $\Phi_{SC}+\Phi_{ECM}=0.1$. Since the porosity directly depends on cell locations, we therefore update $p$ on the {\it fast} 
timescale (and hence $\Phi_{cell}$, $\Phi_{SC}$ and $\Phi_{ECM}$ are also updated on the same time scale).

The cellular contribution to porosity, $\Phi_{cell}$, depends directly on the cell positions on the computational domain. Since the volume at a grid node corresponds to a region that allows for the overlap of cells and the presence of scaffold and extracellular matrix around the cell, we use $\widehat{\Phi}=0.3$ to define the contribution of the cell to the solid volume fraction. This is further depicted in Fig.~\ref{fig:por}(a) where $\Phi_{cell}=0$ outside of the cellular regions, $\Phi_{cell}=0.3$ where there is a non-overlapping cellular region, and $\Phi_{cell}=0.6$ in the regions where there are two cells overlapping. Additional details on cellular geometry and the grid are in Appendix A.

In the growth process of tissue engineered cartilage, the scaffold degrades over time as the cells produce extracellular matrix \citep{Kock12,Matsiko15,Sengers07}. As in \cite{Wilson02}, the time course of the scaffold volume fraction $\Phi_{SC}$ is 
\begin{equation}
\Phi_{SC}(t)=\left\{
\begin{array}{ll}
\Phi_{SC}^0, & t<\tau\\
\\
\Phi_{SC}^0e^{-k_{SC}(t-\tau)}, \quad & t\geq \tau
\end{array}
\right.
\label{eq:sc_sol}
\end{equation}
where $\Phi_{SC}^0$ is the initial scaffold volume fraction, $k_{SC}$ is the rate of scaffold degradation and $\tau$ is the time delay for scaffold degradation, see Table~\ref{tab:par}. The model assumes a uniform distribution of $\Phi_{SC}$ over the computational domain $\Omega$.
The accumulating ECM solid volume fraction $\Phi_{ECM}$ is described by a logistic growth model \citep{Wilson02},
\begin{equation}
\Phi_{ECM}(t)=\Phi_{ECM}^{SS}\ \left(1-e^{-k_{ECM} \cdot t}\right),
\label{eq:ecm_sol}
\end{equation}
where $\Phi_{ECM}^{SS}$ is the steady state value of $\Phi_{ECM}$, and $k_{ECM}$ is the rate of growth, see Table~\ref{tab:par}. The model assumes uniform growth of $\Phi_{ECM}$ (not related to cell position).\\

\subsection{CA model}

The cellular component is described with an off-lattice {\it cellular automaton} model. Cells are discrete entities described as circles with constant diameter $c_d$ with a surface covering 113 grid points spread around the cell center. At time zero, cells are randomly positioned on the domain $\Omega$ to avoid overlap, and cells are initialized with a random age in the interval $[0,3]$ days. Note that we use the same cell initialization for all simulations and we track the center of the cells ($x$ and $y$ coordinates) in the variable \texttt{loc} (of dimension $n_t\times 2$ where $n_t$ represents the number of cells at time $t$). At each time step of the simulation, a cell's potential movement and capability to divide (reproduce) are determined by a set of rules. The present model includes the phenomena of cellular division, random biased off-grid cellular movement (depending on nutrient concentration and porosity), cell-to-cell contact inhibition, and cell death. In the model, the current status of the cell is monitored with the variable {\texttt{cell\_status}}, see Table~\ref{tab:cell_status}. The value of this variable determines the possible actions for each cell during the numerical simulation. Note that a quiescent cell corresponds to a cell that is not moving and can not go through cell division due to a recent collision with another cell. A cell status of 3 denotes a cell that will soon die and be removed from the simulation.
\begin{table}[t]
	\centering
	\begin{tabular}{cl}
		\toprule
		{\bf Numerical value} & {\bf Cell status} (\texttt{cell\_status})\\
		\midrule
		$0$ & moving \\
		$1$ & dividing \\
		$2$ & quiescent \\
		$3$ & dying\\
		\bottomrule
	\end{tabular}
	\caption{Each cell is assigned a status with an associated numerical value, which determines the rules that are applied to that particular cell at a given time step in the \textit{CA} model. }
	\label{tab:cell_status}
\end{table}

\subsubsection{Cellular movement}
\begin{figure}[t!]
	\centering
	\begin{tikzpicture}
	
	\foreach \x in {0,1,...,7}{
		\draw[domain=0:2,smooth,variable=\y] plot ({1*\y*cos(22.5*(2*\x+1))},{1*\y*sin(22.5*(2*\x+1))});
	}
	\draw[very thick,fill=gray!50] (0,0) circle (0.5cm);
	\foreach \x in {-3,-2.75,...,3}{
		\foreach \y in {-3,-2.75,...,3}{
			\draw[very thick] (\x+0.2,\y+0.1) circle (0.005cm);
	}}
	\foreach \x in {-2.875,-2.625,...,2.875}{
		\draw[thin] (\x+0.2,-3+0.1) -- (\x+0.2,3+0.1);}
	\foreach \y in {-2.875,-2.625,...,2.875}{
		\draw[thin] (-3+0.2,\y+0.1) -- (3+0.2,\y+0.1);}
	\draw [very thick] (0,0) circle (2cm);
	\node[right,blue](E) at (2,0) {\bf E};
	\node[above right,blue](NE) at (1.414,1.414) {\bf NE};
	\node[above,blue](N) at (0,2) {\bf N};
	\node[above left,blue](NW) at (-1.414,1.414) {\bf NW};
	\node[left,blue](W) at (-2,0) {\bf W};
	\node[below left,blue](SW) at (-1.414,-1.414) {SW};
	\node[below,blue](S) at (0,-2) {\bf S};
	\node[below right,blue](SE) at (1.414,-1.414) {SE};
	\end{tikzpicture}
	\caption{Schematic of a cell (shaded gray circle in center) on the computational domain where the eight cardinal and intercardinal directions ($E$, $NE$, $N$, $NW$, $W$, $SW$, $S$, $SE$) are shown. The circle on each grid box corresponds to the node value tracking the average porosity and nutrient concentration on that particular volume element. The nutrient and porosity values are assessed in the direction of no movement $\mathbf{\varnothing}$, on the nodes included in the shaded gray area, and in the eight cardinal and intercardinal directions, on the nodes contained in the annulus surrounding the cell. The inner diameter of the annulus corresponds to $c_d$ while the outer diameter of the annulus $a_{d,mov}$ is reported in Table~\ref{tab:par}.}
	\label{fig:cell-nbhd}
\end{figure}
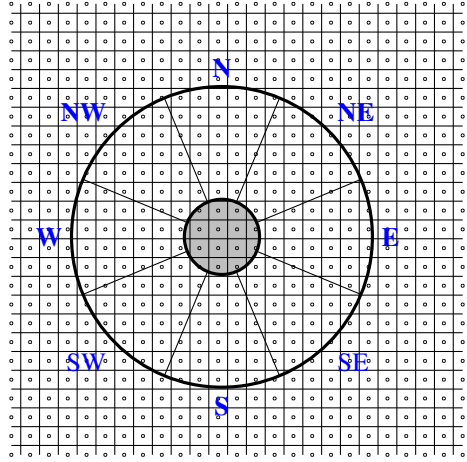

Cellular movement is random and it is biased by nutrient concentration (chemotaxis) and porosity \citep{Erickson09,Matsiko15,Morales07,Freyria04}. For each cell that is in the moving state (\texttt{cell\_status=0}), as shown in Fig.~\ref{fig:cell-nbhd}, the cell will move in nine possible directions denoted by: $\mathbf{\varnothing}$, $E$, $NE$, $N$, $NW$, $W$, $SW$, $S$, $SE$. The first direction corresponds to the small probability of staying in the same exact position if the cell is in the moving state whereas the other directions correspond to the cardinal and intercardinal directions.

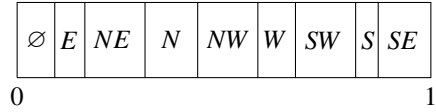
\begin{figure}[th]
	\centering
	\begin{tikzpicture}
	\draw node[below]{0}(0,0) -- (5.5,0) node[below]{1} -- (5.5,1) -- (0,1) --node[right,pos=0.5]{$\mathbf{\varnothing}$} (0,0);
	\draw (0.5,0) -- node[right,pos=0.5,xshift=-0.05cm]{$E$} (0.5,1);
	\draw (0.9,0) -- node[right,pos=0.5]{$NE$} (0.9,1);
	\draw (1.7,0) -- node[right,pos=0.5,xshift=0.1cm]{$N$} (1.7,1);
	\draw (2.4,0) -- node[right,pos=0.5]{$NW$} (2.4,1);
	\draw (3.2,0) -- node[right,pos=0.5,xshift=-0.05cm]{$W$} (3.2,1);
	\draw (3.7,0) -- node[right,pos=0.5]{$SW$} (3.7,1);
	\draw (4.5,0) -- node[right,pos=0.5,xshift=-0.05cm]{$S$} (4.5,1);
	\draw (4.8,0) -- node[right,pos=0.5]{$SE$} (4.8,1);
	\end{tikzpicture}
	\caption{Example of direction vector for cell movement. The width of each bin is computed following \eqref{eq:omegan}, \eqref{eq:omegap}, and \eqref{eq:omega} for the direction vector \texttt{dir\_n}, \texttt{dir\_p}, and \texttt{dir}, respectively.}
	\label{fig:dir}	
\end{figure}

In order to determine the cell's movement direction, 
the average nutrient concentration $c_{avg,j}$ is calculated on the nodes of the computational grid covered by the cell (direction $j=\mathbf{\varnothing}$) and in the eight neighborhoods exterior to the cell, as depicted in Fig.~\ref{fig:cell-nbhd}. The average values of nutrient concentration are then normalized in a probability direction vector where higher nutrient concentration will correspond to a higher probability in the {\it nutrient direction vector}, \texttt{dir\_n}, shown in Fig.~\ref{fig:dir}. The width of each bin $\omega_{n,j}$, in Fig.~\ref{fig:dir}, corresponds to the probability of moving in each of the directions and is computed as 
\begin{equation}
\omega_{n,j}=\displaystyle \dfrac{c_{avg,j}}{\sum_{i=\mathbf{\varnothing}}^{SE} c_{avg,i}} \qquad j=\mathbf{\varnothing},E,\ldots,SE.
\label{eq:omegan}
\end{equation}
The nutrient direction vector \texttt{dir\_n} is obtained by sequentially combining rescaled $\widehat{\omega}_{n,j}$ bins in a probability vector that is in the range $[0,1]$. Similarly, a second direction vector is obtained for the average porosity where higher porosity will correspond to a higher probability in the {\it porosity direction vector}, \texttt{dir\_p}. The width of the bins $\omega_{p,j}$ of \texttt{dir\_p} are computed as follows
\begin{equation}
\omega_{p,j}=\displaystyle \dfrac{p_{avg,j}}{\sum_{i=\mathbf{\varnothing}}^{SE} p_{avg,i}} \qquad j=\mathbf{\varnothing},E,\ldots,SE
\label{eq:omegap}
\end{equation}
where the quantity $p_{avg,j}$ represents the average porosity on the grid points in the region $j$ (i.e. fluid region or $1-\Phi$ for solid volume fraction $\Phi$). The width of a bin is set to zero if the corresponding region contains a grid point with porosity less than $p_{sens}$,  which corresponds to an unfavorable region since there is not enough room for the cell. The bin widths are rescaled as $\widehat{\omega}_{p,j}$ to again ensure a range of $[0,1]$. Finally, the two direction vectors are combined to obtain a comprehensive direction vector, \texttt{dir}, for each cell. The width of the bins, $\omega_j$, of the vector \texttt{dir} are obtained as
\begin{equation}
\omega_j=\widehat{\omega}_{n,j} \cdot\widehat{\omega}_{p,j} \qquad j=\mathbf{\varnothing},\ldots,SE
\label{eq:omega}
\end{equation}
In order for the bins to be in the range of $[0,1]$, the bins $\omega_j$ are rescaled as $\widehat{\omega}_j$ and are then  sequentially combined in the direction vector \texttt{dir} as in Fig.~\ref{fig:dir}. The movement direction of each cell is determined by choosing a distinct random variable \texttt{X} from a uniform distribution $U[0,1]$ and determining which direction bin contains $\texttt{X}$. We note that wide bins $\widehat{\omega}_j$ have a higher probability of containing \texttt{X} and have larger width due to a more favorable nutrient concentration (larger $c_{avg,j}$) and/or a more favorable porosity (larger $p_{avg,j}$). As an example of how the direction is chosen, if \texttt{X}=$0.35$ and the rescaled bin sizes are $\widehat{\omega}_{\mathbf{\varnothing}}=0.1$, $\widehat{\omega}_E=0.1$, and $\widehat{\omega}_{NE}=0.2$, \texttt{X} will fall in the $NE$ bin (since $0.1+0.1+0.2=0.4$), therefore the corresponding cell will move in the $NE$ direction. We use the convention that the chosen cell direction $k$ is the first direction to satisfy the following inequality: \texttt{X}$<\left(\sum_{j=\varnothing}^{k}\widehat{\omega}_j\right)$ for $k={\mathbf{\varnothing}},E,\ldots,SE$. We note that if \texttt{X}$=1$, this will correspond to $k=SE$. At a given time step, all cells will simultaneously move in the chosen direction $k$ a fixed distance $d_{cell}$ (reported in Table~\ref{tab:par}). The movement is applied to the cell center coordinates stored in the matrix \texttt{loc}. Again, if $NE$ is chosen for cell $i$, then we update the x and y coordinates of the $i^{th}$ cell by adding $d_{cell}\cos(\pi/4)$ and $d_{cell}\sin(\pi/4)$ to the old locations, respectively. Note that if cell movement results in a local porosity with $p>1$, which may occur on occasion when multiple cells have moved into the same location, this issue is resolved by randomly choosing to undo the movement of half of these cells, until the situation is resolved.

%
Movement of cells may result in collisions among cells, which will cause a cell to enter the quiescent state (\texttt{cell\_status}=2). The quiescent state is due to the cell-to-cell contact inhibition  and cells will stay idle in this state (no movement or division) for a fixed time $t_{idle}$, reported in Table~\ref{tab:par}. At the end of each iteration, we check for new cell collisions by determining which cells in the moving state (\texttt{cell\_status}=0) are  now a distance less than $c_d$ apart from any other cell (distance between cell centers). The cells that are less than $c_d$ apart will enter the quiescent state (\texttt{cell\_status}=2); the collision counter is initially set to $t_{idle}$ and is then decreased by $\triangle t$ at each iteration. When the counter reaches 0, the cell is then put back into the moving state (\texttt{cell\_status}=0). 

\subsubsection{Cellular division and death}
As shown in the algorithm diagram in Fig.~\ref{fig:flux-big}, at each time step, we first check whether a cell will be tagged to die in this time step and whether a cell has the potential to divide. These decisions will be based on the rules outlined below. 

The rule that governs the capability of a cell to undergo cellular division is related to four different conditions that must be simultaneously satisfied at a given time point. In order for cell division to occur, a cell must:
\begin{enumerate}
	\item Not be in the quiescent or dying mode (\texttt{cell\_status}$\neq$2,3).
	\item Have reached cellular maturity (\texttt{cell\_age}$\geq t_{div}$, see Table~\ref{tab:par}).
	\item Be in a region with enough nutrient ($c_{avg} > c_{div}$, see Table~\ref{tab:par}, where $c_{avg}$ is the average nutrient concentration on the grid points covered by the cell surface).
	\item Be nearby a region with high enough porosity to host a new cell ($p>p_{sens}$, see Table~\ref{tab:par}).
\end{enumerate}
The cells that have the potential to divide (passing all necessary conditions listed above) undergo an overcrowding check to ensure that there is room for the new daughter cell. The center-to-center distance between all dividing cells is computed.
Of all the cells that are less than $d_{div}$ apart, half are chosen at random and are allowed to undergo cell division (and moved to the dividing status, \texttt{cell\_status}=1) where the two daughter cells are each the same size as the mother cell. This check prevents too many cells that are close to each other from dividing during the same iteration. The new cell is placed adjacent to the mother cell along a direction that is randomly chosen depending on the local value of porosity. The procedure used to determine the division direction vector, \texttt{dir\_d}, is similar to the one followed to generate \texttt{dir\_p}. The number of possible directions is reduced from 8 to 6 ($E$, $NE$, $NW$, $W$, $SW$, $SE$, each with an arc of $\pi/3$), so that each of the six regions are able to contain the full surface of the new cell (additional details in Appendix A). The width of the six bins, $\omega_{d,j}$, in the vector \texttt{dir\_d} are
\begin{equation}
\omega_{d,j}=\displaystyle \dfrac{\exp(100p_{avg,j})-1}{\sum_{i=E}^{SE} \exp(100p_{avg,i})-1} \qquad j=E,NE,\ldots,SE,
\label{eq:omegad}
\end{equation}
and the width of a bin is set to zero if the corresponding region contains a grid point with porosity such that $p<p_{sens}$. Similarly, we follow the same procedure used for cell movement where we determine a rescaled $\widehat{\omega}_{d,j}$ such that we remain on $[0,1]$ and we then choose a uniform variable \texttt{X} for each mother cell and find the bin and direction that will determine the position of the new (daughter) cell. The region used to determine the vector \texttt{dir\_d}, similar to the one depicted in Fig.~\ref{fig:cell-nbhd}, has outer diameter $a_{d,div}$, reported in Table~\ref{tab:par}. The new daughter cell will be located at a radial distance of $7c_d/6$ in the chosen direction where the 7/6 ensures that these cells are close but will not automatically go into the quiescent state. To avoid consecutive divisions of the same cells, after a cell divides, the variable \texttt{cell\_age} of the two daughter cells is reset to zero. The variable \texttt{cell\_age} is then incremented by $\triangle t$ after each time step to track the aging process of the cell. At the end of a time step, all cells in the dividing state (\texttt{cell\_status=1}) are then placed into the moving state (\texttt{cell\_status=0}) and then re-evaluated at the next time step. We emphasize that Equation~\eqref{eq:omegad} provides a stronger bias than~\eqref{eq:omegap} resulting in a high probability that the new cell will be placed in a region of high porosity.

Cell death is known to be a random process and we assume there is a small probability of a cell switching on the apoptotic pathway to initiate cell death at each time step. In particular, staining for dead mesenchymal stem cells (precursors to chondrocytes) seeded in gels of different macromer concentrations showed a uniform distribution of dead cells throughout the construct at day 21 \citep{Erickson09}. We convert cellular death rates from continuous models \citep{Bandeiras15} and obtain a death probability per iteration as $p_{death}=0.0006$, reported in Table~\ref{tab:par}. In simulations, at the beginning of the time step, we randomly choose cells to be moved to the dying state (\texttt{cell\_status}=3), where no movement will occur during the current time step, and they will be removed from the domain at the beginning of the next iteration before the dying cells for the next iteration are chosen.

\subsubsection{Boundary Conditions}
In Fig.~\ref{fig:sc}(b), the computational domain $\Omega$ is highlighted in dark gray. Since cells tend to stay within the bio-construct, we assume a no-flux boundary condition on the right and bottom that are exposed to the nutrient bath. The left and top are connected to the rest of the domain and we also assume a no-flux boundary condition since it is assumed that on average, if one cell was leaving this region, another would be entering. If the chosen direction would place a cell outside of the domain, its trajectory is corrected before the cell moves to keep the cell inside the domain. The correction is performed on the coordinate that would exceed the computational domain by reversing the chosen movement direction and reducing its magnitude by a factor of ten. The boundary of the computational domain is padded with several layers of ghost grid points initialized with a value of $c=0$ for concentration and $p=0$ for porosity. Therefore, when a cell is close to the boundary, if part of the region depicted in Fig.~\ref{fig:cell-nbhd} covers any ghost node, that specific direction will be automatically penalized by the direction rules \eqref{eq:omegan}, \eqref{eq:omegap} and \eqref{eq:omegad} to minimize the chances that a cell will exit the domain or will divide outside the domain. 

\section{Results and Discussion}
A primary aim of this study was to incorporate the influence of scaffold porosity, and hence initial scaffold solid volume fraction, to then investigate cellular movement and reproduction in a cell-seeded scaffold with applications to articular cartilage regeneration. To validate our model and identify parameter values that led to tissue growth, we first started with data from \cite{Freed94} where the chondrocyte density was reported for polyglycolic acid (PGA) scaffolds with initial scaffold porosities of 0.92-0.96 ($\Phi_{SC}^0\in$[0.04,0.08]) with different cell seedings. We extracted their cell density data, which was reported as an average over three samples  for seven different experiments, where scaffold dimension and initial cell density was varied. To obtain the number of cells to seed in our simulation domain $\Omega$, we found the product of our domain volume and the mean cell seeding density (cells/cm$^{3}$) of the seven different experiments, which is equal to $2000$ cells. We then reduced this since experiments of \cite{Freed94} reported an elevated death rate in the first $2$ days (~$50\%$) and we use a constant death probability over the course of the simulation.

\begin{figure}[ht]
	\begin{center}
		\includegraphics[width=0.95\textwidth]{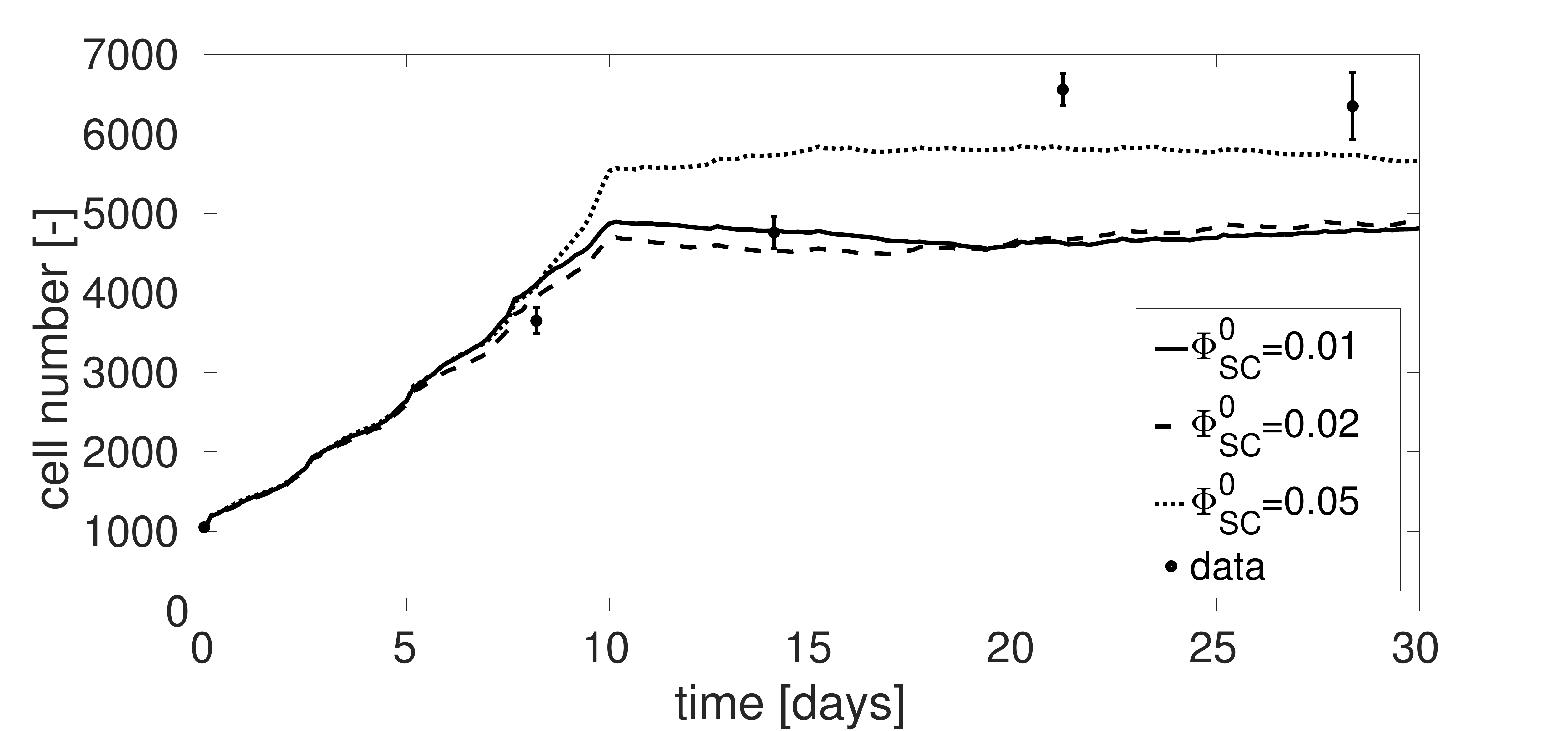}
	\end{center}
	\caption{Total cell counts in the bio-construct with respect to time for the baseline parameters given in Table \ref{tab:par} and using different values of initial scaffold macromer concentrations of 1\%, 2\%, and 5\%, which correspond to initial scaffold volume fractions of $\Phi^0_{SC}=0.01$, 0.02, and 0.05, respectively. The model predictions are compared with chondrocytes seeded in a polygycolic acid (PGA) scaffold from~\cite{Freed94}, shown as data points representing the mean and one standard deviation over three samples. 
	}
	\label{fig:comp-freed}
\end{figure}

Using the parameters reported in Table \ref{tab:par}, we start with 1000 cells in our domain $\Omega$ and track the total number of cells through 30 days. These results are shown in Fig.~\ref{fig:comp-freed} for three different initial scaffold volume fractions ($\Phi^0_{SC}=0.01$, $0.02$, and $0.05$). For comparison, we also plot on this graph the data of \cite{Freed94} at day 0, 8, 14, 21, and 28  with circles denoting the mean and bars showing one standard deviation across three samples of scaffold thickness $h=0.168$ cm and diameter $d=1$ cm, with initial seeding in the total construct of $2\cdot 10^6$ cells. Note that the Freed data is actually for days 2, 10, 16, 23, and 30 of their experiment but since we start with the cell numbers reported at day 2, we have shifted their data to compare with the simulations. The slope of the cell count curves from day 0 to 8 are all in good agreement with the slope of the experimental data points. Overall, we find relatively good agreement between this data and our simulations; the cell densities of our simulations are slightly smaller than the experiment at later time points for this set of baseline parameters. For the first 8 days, all three simulation cases result in similar cell growth and after day 8, the scaffold with 5\% initial scaffold concentration (i.e. $\Phi^0_{SC}=0.05$ ) starts to significantly increase in cell number, leveling off at around 5800 cells. The 5\% case has the highest cell density, closer to the cell counts of \cite{Freed94} where initial scaffold concentrations were 4-8\%. For the case of  $\Phi_{SC}^0=0.08$ with baseline parameters, this bio-construct achieves a maximum cell count of 8000 cells and a final cell count of 6900 cells, showing  good agreement with the data from~\cite{Freed94} (results not shown). 

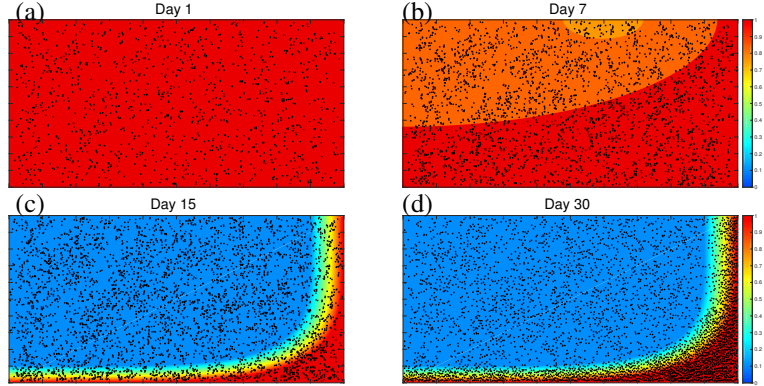
\begin{figure}[ht]
	\begin{center}
		\begin{tikzpicture}
		\node (fig) at (0,0) {\includegraphics[width=0.95\textwidth]{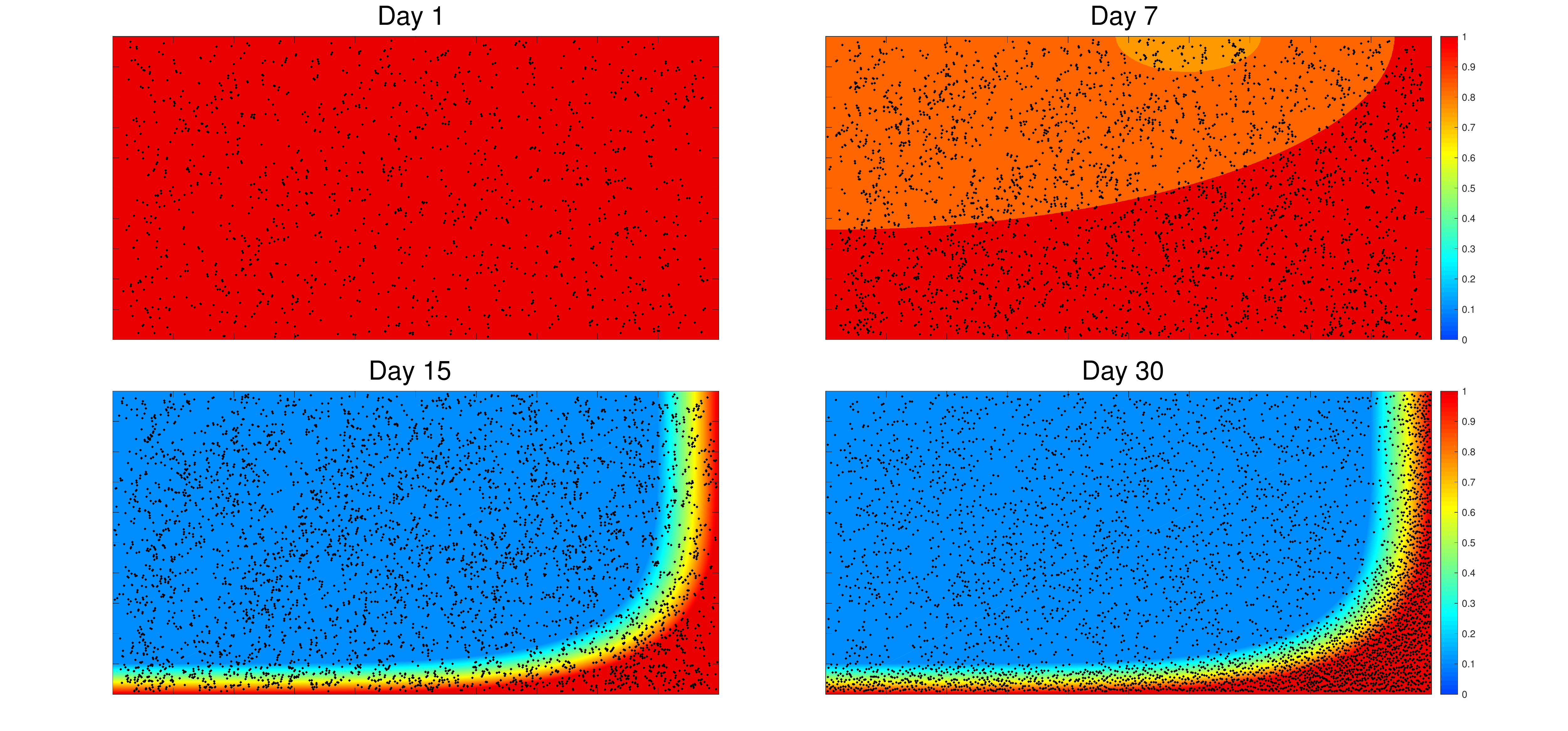}};
		\node (a) at (-4.65,2.5) {(a)};
		\node (b) at (0.5,2.5) {(b)};
		\node (c) at (-4.65,-0.05) {(c)};
		\node (d) at (0.5,-0.05) {(d)};
		\end{tikzpicture}
	\end{center}
	\caption{Cell distribution for the baseline scenario using parameters in Table \ref{tab:par} with an initial scaffold volume fraction of 1\% ($\Phi_{SC}^0=0.01$) at day 1 in (a), day 7 in (b), day 15 in (c), and day 30 in (d). The non-dimensional nutrient profile is shown in the background with the colorbars on the right. Cell locations are denoted with black dots on the domain. Note that the minimum value of the nutrient profile in the upper left corner of panel (d) is $0.1$.}
	\label{fig:nutrient}
\end{figure}

To highlight the cellular dynamics in the evolving bio-construct, in Fig.~\ref{fig:nutrient}, the cell locations in the construct are plotted at four time points for the case of a scaffold with $\Phi_{SC}^0=0.01$. For all simulations, we assume the same nutrient profile as detailed in Appendix B where at day 1, Fig.~\ref{fig:nutrient}(a), there is a constant and high nutrient concentration throughout the construct. Thus, at this time point, cells will primarily move in directions corresponding to higher porosity (more free space due to less cells in the region). At later time points such as day 7, Fig.~\ref{fig:nutrient}(b), due to cell division in the interior of the construct, the nutrient concentration is decreasing in these regions. Cell motility will be biased now based on nutrient concentration and porosity. The bias of cells to migrate to regions of higher nutrient concentration is clearly seen at time points of day 15, Fig.~\ref{fig:nutrient}(c), and day 30, Fig.~\ref{fig:nutrient}(d), where the nutrient profile continues to decrease on the interior portions of the construct and only remains high in the regions closest to the nutrient bath. At day 30, a band of cells at a higher density is clearly seen and we still observe cells in the upper left corner since due to porosity constraints, there is not enough room for all cells to migrate to the boundaries. We note that similar profiles of cellular distributions are observed for other cases of initial scaffold solid volume fractions. The simulation results also match up well with previous experiments of \cite{Vunjak98} and models of \cite{Chung10} in terms of increased cellular distributions near edges with higher nutrient concentration. However, in comparison to the model of \cite{Chung10}, the dynamics of aggregation are a bit different since we are also accounting for a porosity bias in movement and cellular division. The local porosity is changing in time due to the local cell volume fraction as well as the decreasing scaffold volume fraction and increasing ECM volume fraction.

The results shown in Fig.~\ref{fig:comp-freed} and \ref{fig:nutrient} are for a single run of the model. However, the hybrid modeling framework we utilize involves several terms that are stochastic and hence there is aleatory uncertainty~\citep{Alden13,Cosgrove15,Read12} in model results for a single simulation. Since we want to ensure that the simulation results account for the full amount of variation in the system, we utilize an \textit{A-test}~\citep{Vargha00} as described in Appendix C. The \textit{A-test} determines the number of simulations to run to ensure that the results we report are a true representation of the entire model. For all simulation results in the remaining figures, the reported means and standard deviations correspond to 300 simulations with the same cell initial location and age. 

\begin{figure}[ht]
	\begin{center}
		\begin{tikzpicture}
		\node (fig) at (0,0) {\includegraphics[width=0.95\textwidth]{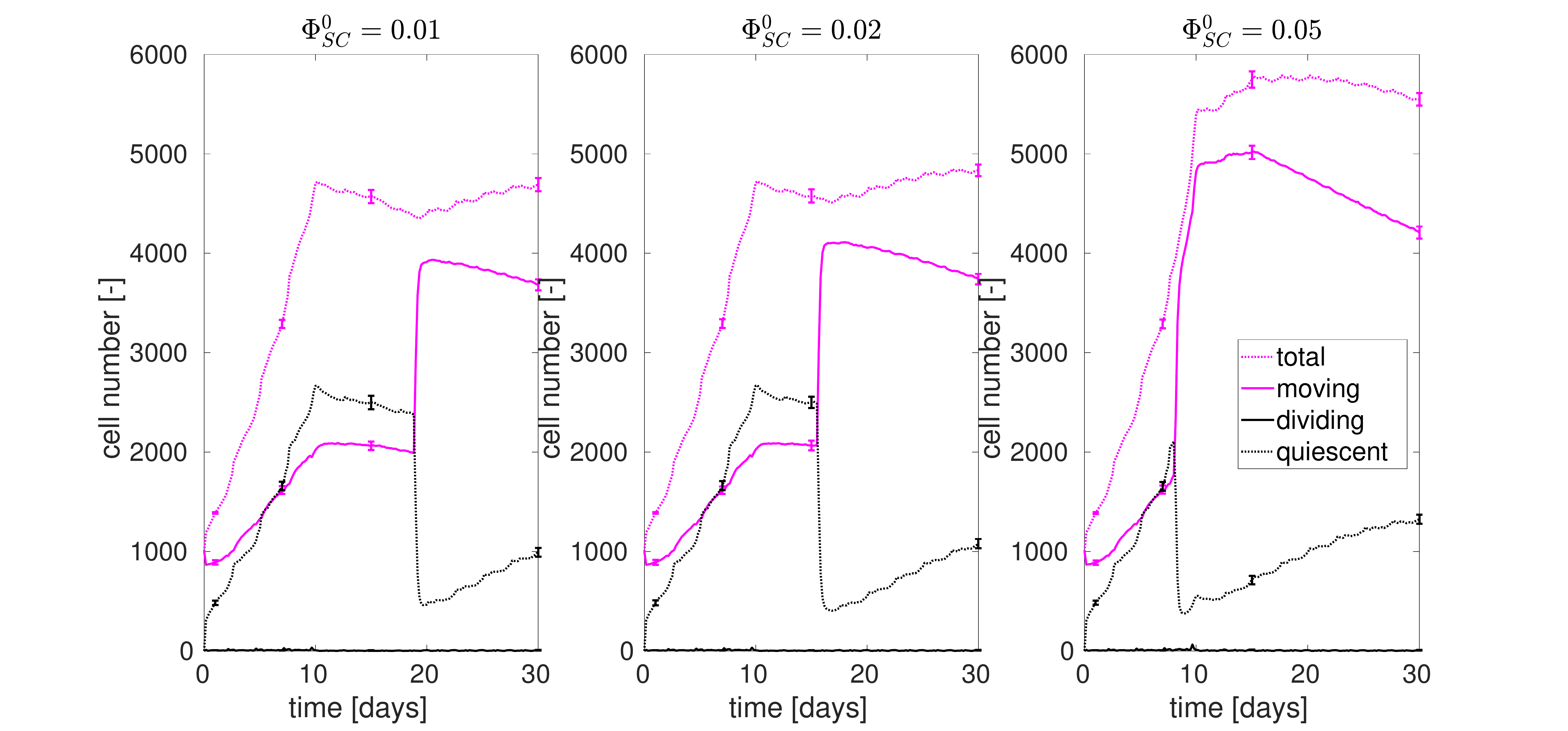}};
		\node (a) at (-4.5,2.5) {(a)};
		\node (b) at (-1.35,2.5) {(b)};
		\node (c) at (1.85,2.5) {(c)};
		\end{tikzpicture}
	\end{center}
	\caption{Model predictions of cells in different states for the baseline parameters given in Table \ref{tab:par} with different values of scaffold volume fractions of $\Phi^0_{SC}=0.01$, 0.02, and 0.05, respectively. The graph displays the total number of cells (dotted magenta), the cells in the moving state (solid magenta), cells in the dividing state (solid black), and cells in the quiescent state (dotted black). Curves and bars correspond to the mean and standard deviation over 300 simulations.}
	\label{fig:baseline}
\end{figure}
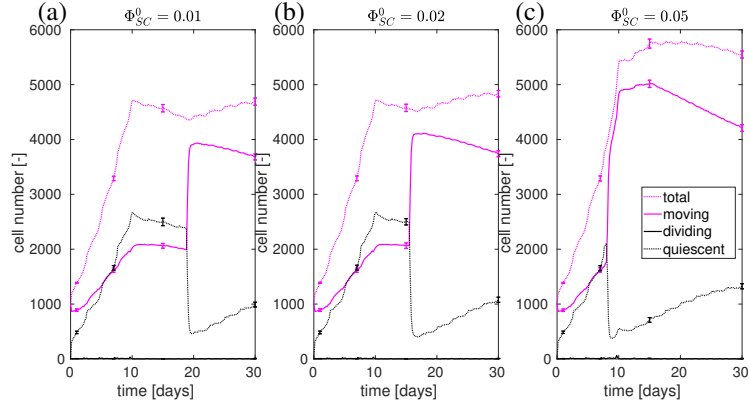

To illustrate the dynamics of cells for the baseline case, we track  cell counts for each of the different cell states in Fig.~\ref{fig:baseline}. We emphasize again that the nutrient profile and ECM accumulation is assumed to be the same in each case. The only parameter that is changed is the initial scaffold volume fraction $\Phi_{SC}^0$, which will then degrade with similar dynamics after a time delay of $\tau$ days. Similar behavior in terms of the total cell counts in the different states are observed for $\Phi_{SC}^0=0.01$ and 0.02. One notable difference is that the number of moving cells increases significantly around day 20 for the 0.01 case, Fig.~\ref{fig:baseline}(a), and increases around day 14 for the 0.02 case, Fig.~\ref{fig:baseline}(b). This trend results in a similar decrease in  quiescent cells at the same time points. This can be partially explained as follows. Cells initially in the highest porosity (lower volume fraction of scaffold) are more free to move within the domain at earlier time points. As the nutrient profile changes, this then biases cells to move towards nutrient rich regions and will often result in cells colliding with each other. Once cells adjust their movement again, they are able to bias their motion in directions balancing higher nutrient and higher porosity (and hence, less cells nearby). In the case of lower initial porosity (higher initial scaffold volume fraction), cells are more biased by porosity initially and are able to move throughout the domain with less collisions. Hence, they are able to distribute themselves in a way that allows space for new daughter cells. Note that we are accounting for cell death in these simulations; this is on the order of 25-100 cells per day, which is why these curves are observed as very small oscillations. 

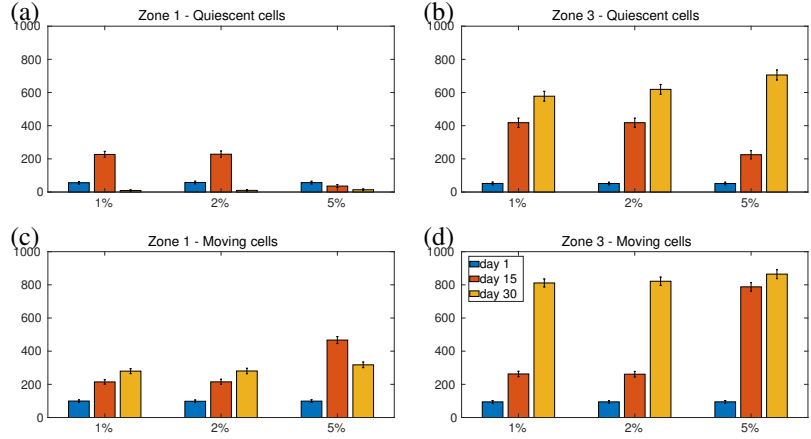
\begin{figure}[ht]
	\begin{center}
		\begin{tikzpicture}
		\node (z1q) at (0,0) {\includegraphics[width=0.49\textwidth]{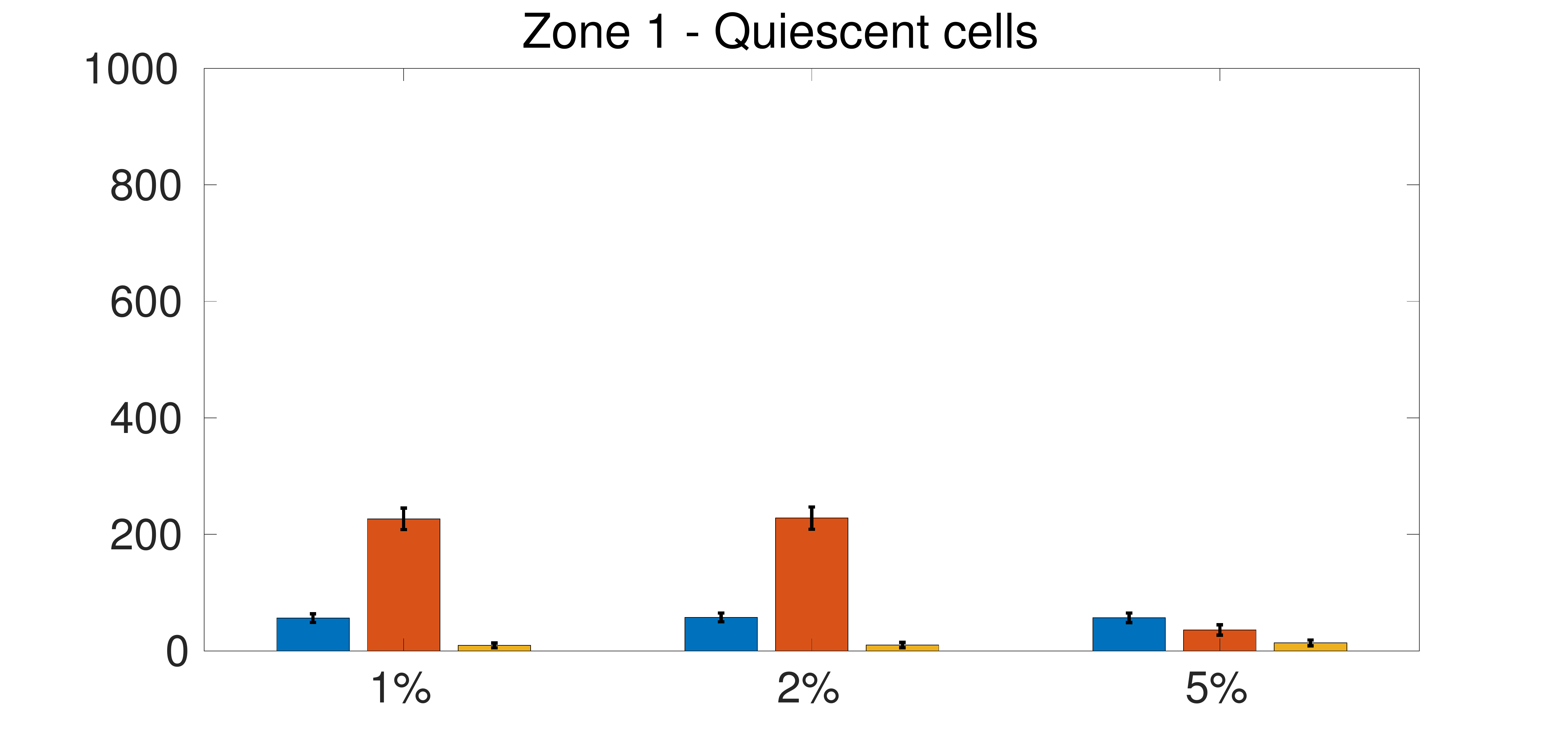}};
		\node (z3q) at (5.5,0) {\includegraphics[width=0.49\textwidth]{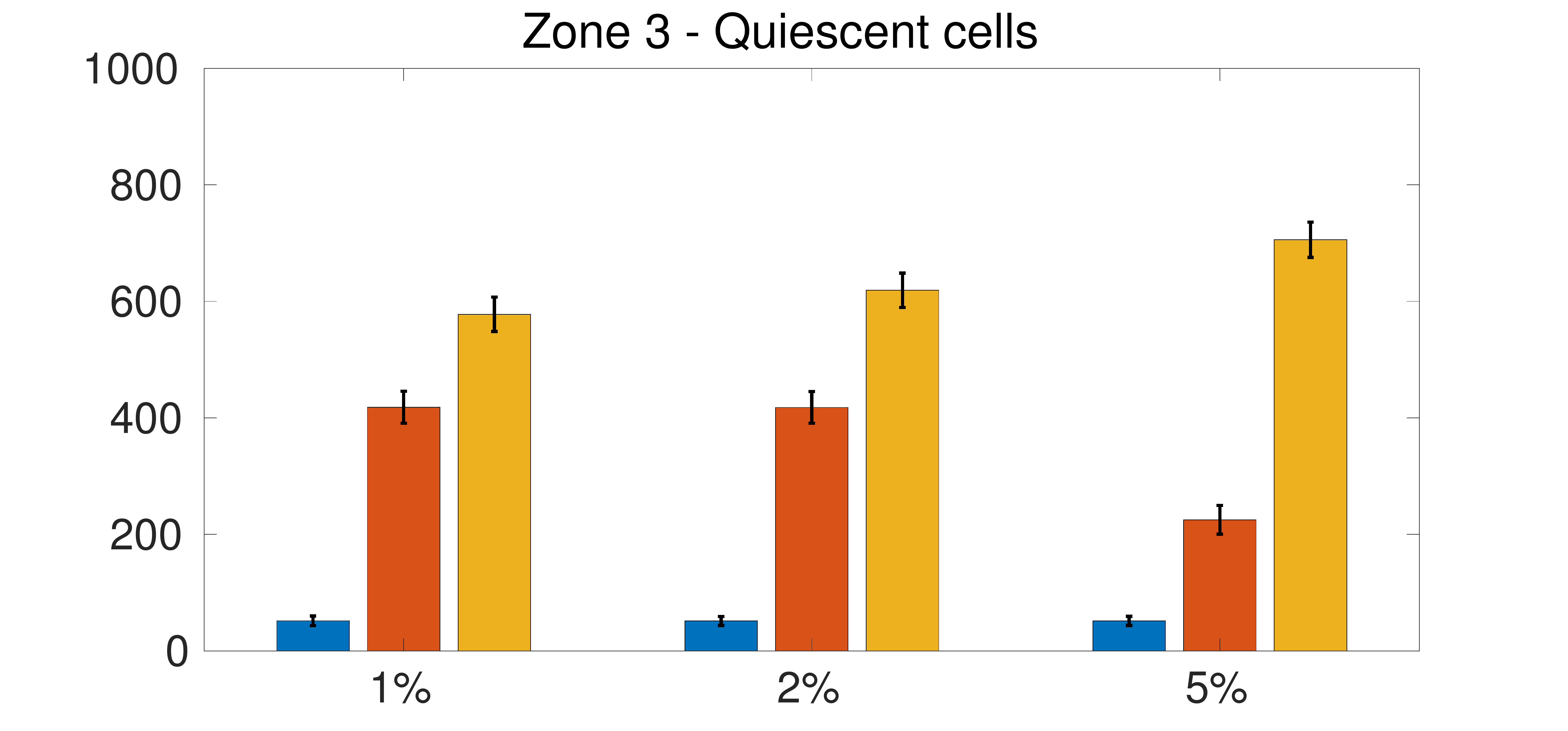}};
		\node (z1m) at (0,-3) {\includegraphics[width=0.49\textwidth]{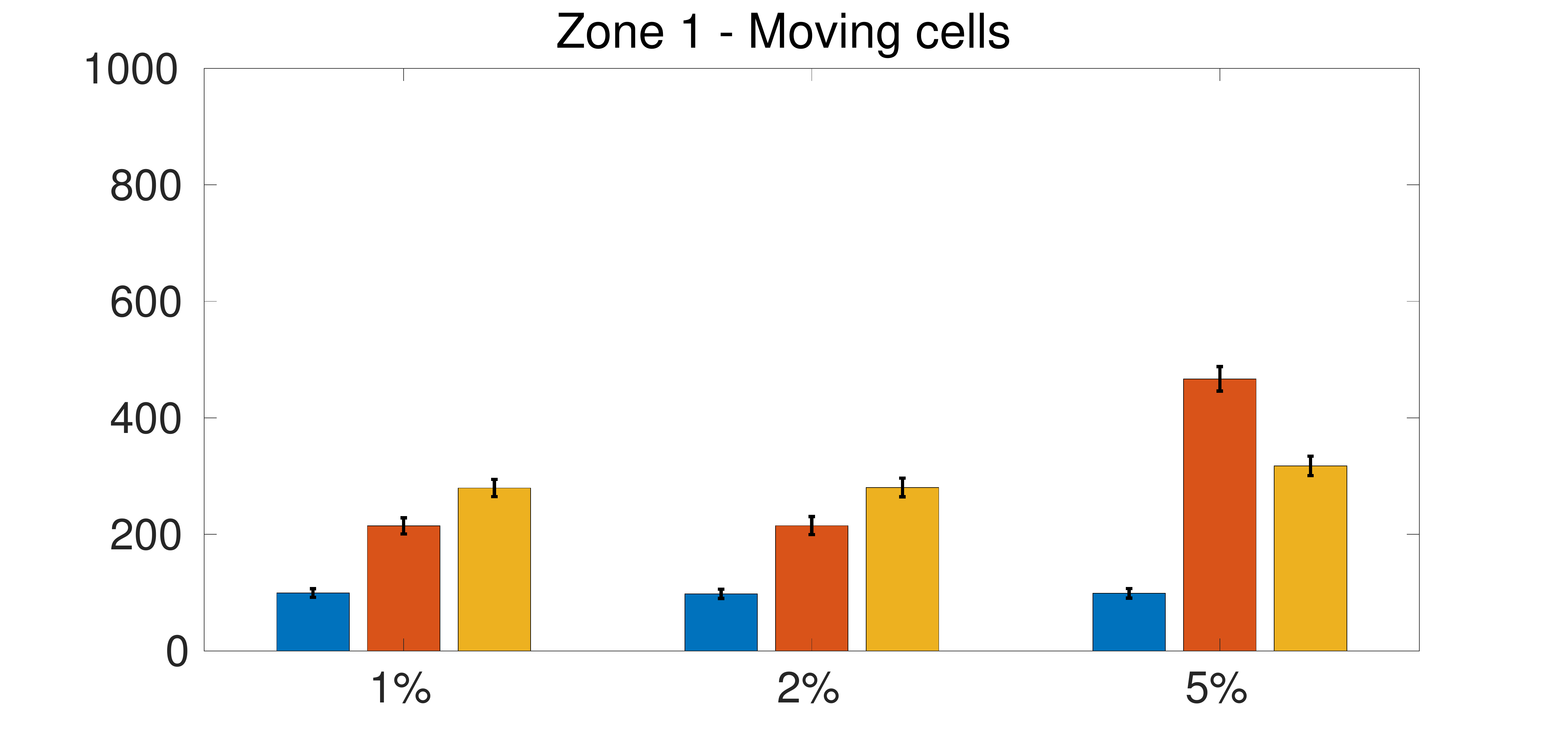}};
		\node (z3m) at (5.5,-3) {\includegraphics[width=0.49\textwidth]{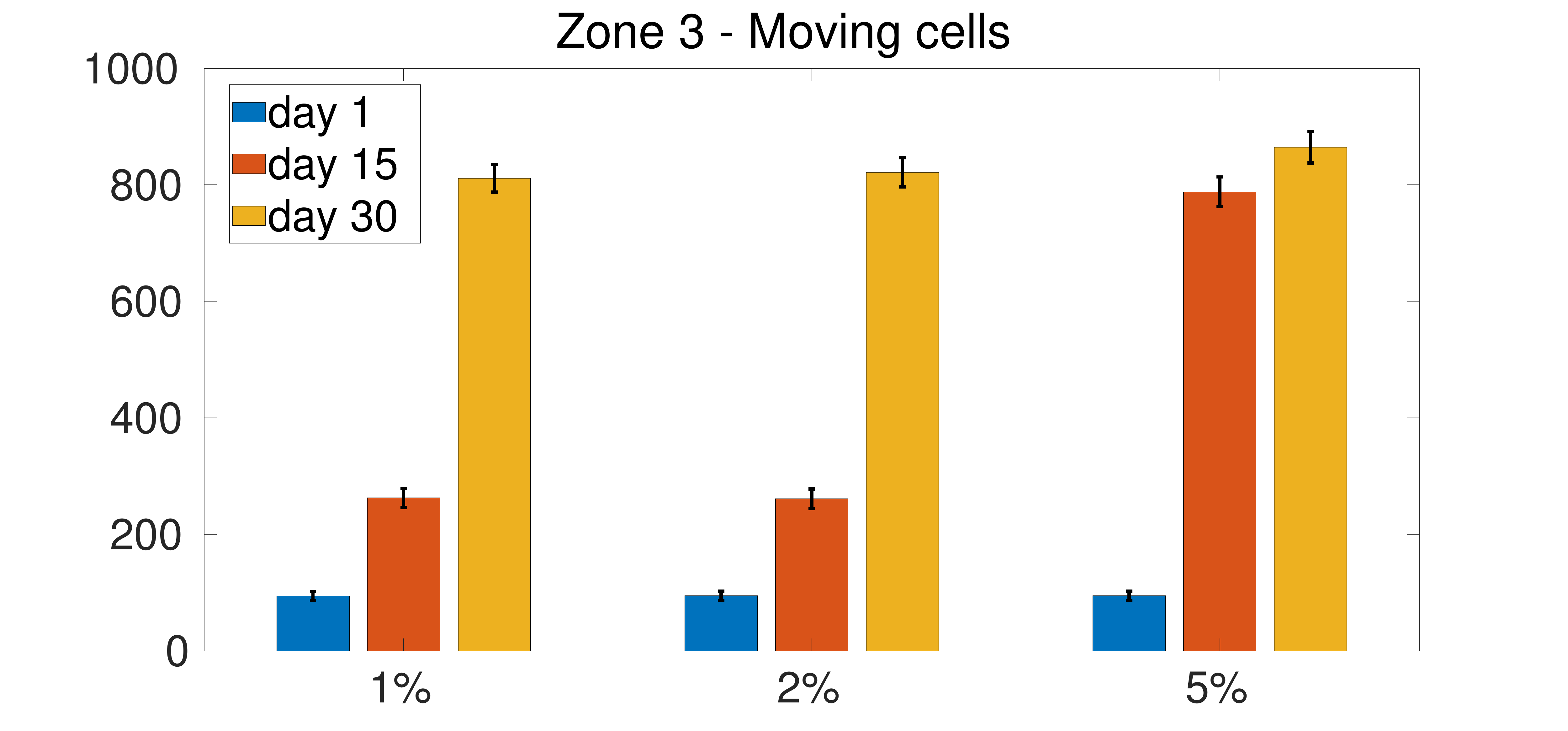}};
		\node (a) at (-2.5,1.3) {(a)};
		\node (b) at (3,1.3) {(b)};
		\node (c) at (-2.5,-1.7) {(c)};
		\node (d) at (3,-1.7) {(d)};
		\end{tikzpicture}
	\end{center}	
	\caption{Cell counts by cellular status in Zone 1 and Zone 3, the upper left and lower right corners of the domain $\Omega$ shown in Fig.~\ref{fig:sc}(b), for $\Phi_{SC}^0=0.01$ (1\%), $\Phi_{SC}^0=0.02$ (2\%), and $\Phi_{SC}^0=0.05$ (5\%) at day 1, 15, and 30. Quiescent cell counts are shown in the top row with (a) Zone 1 and (b) Zone 3. Moving cell counts are shown in the bottom row with (c) Zone 1 and (d) Zone 3. Simulation parameters are reported in Table~\ref{tab:par}.}
	\label{fig:barbaseline}
\end{figure}

As observed in Fig.~\ref{fig:nutrient}, we observe clusters of cells closer to the nutrient rich regions at day 14 and later time points. To further characterize these spatial differences, we analyze the cell status of cells  in two regions of the domain $\Omega$ shown in Fig.~\ref{fig:sc}(b). We refer to the upper left corner of the domain, $[0,0.17]\times[0,0.03]$, as Zone 1 and the lower right corner of the domain $[0.33,0.5]\times[0.07,0.1]$,  as Zone 3. In Fig.~\ref{fig:barbaseline}, we highlight the differences in quiescent and moving cells at day 1, 15, and 30. For the quiescent cells, we observe different trends in Zone 1 and Zone 3, shown in Fig.~\ref{fig:barbaseline}(a)-(b). In Zone 3, we observe a similar monotonic increasing behavior as days increase for all three initial scaffold macromer concentrations. This is most likely due to higher nutrients in Zone 3, causing increased collisions due to increased cells in this Zone. In contrast, there is an increase and then decrease in quiescent cells for the 1 and 2\% cases whereas the 5\% decreases with time. For the 5\% case, this is due to the fact that in this model, a lower initial porosity with the same porosity sensitivity means that on average, these cells will be moving smarter in the direction of higher nutrient, hence moving out of Zone 1.  The counts for moving cells are shown in Fig.~\ref{fig:barbaseline}(c)-(d), where Zone 1 in (c) has a similar monotonic increasing trend for the 1 and 2\% cases. However, the 5\% case increases and then decreases, due to cells moving out of Zone 1 and towards the nutrient rich regions. The number of moving cells in Zone 3 is shown in (d), where we observe a monotonic increasing trend with for all three cases. 
We note that for the total cell counts (results not shown) in Zone 1 we observe, for all three cases, an increase from day 1 to 15 whereas there is a decrease from day 15 to 30 due to the decreased nutrient concentration at later time points. All three cases also have approximately the same number of total cells in Zone 1; the low nutrient concentration is dominating over porosity in this region at later time points. In Zone 3, close to the nutrient-rich bath, all three cases exhibit an increase in total cell number from day 1 to 15 to 30. The 1 and 2\% scaffold concentrations have less total cells in Zone 3 relative to the 5\% case, similar to results shown for the entire construct in Fig.~\ref{fig:baseline}.

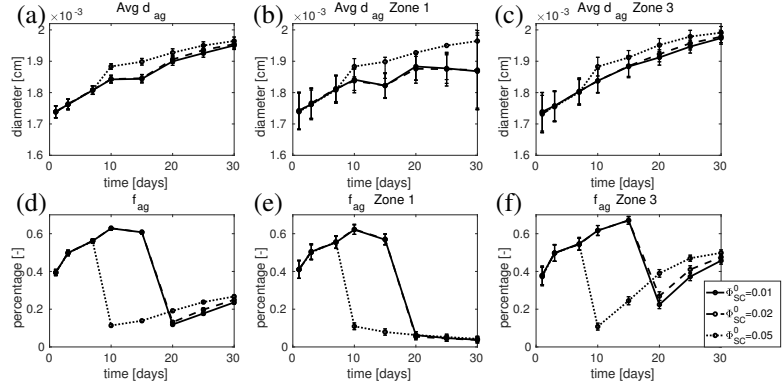
\begin{figure}[ht]
	\begin{center}
		\begin{tikzpicture}
		\node (fig) at (0,0) {	\includegraphics[width=0.95\textwidth]{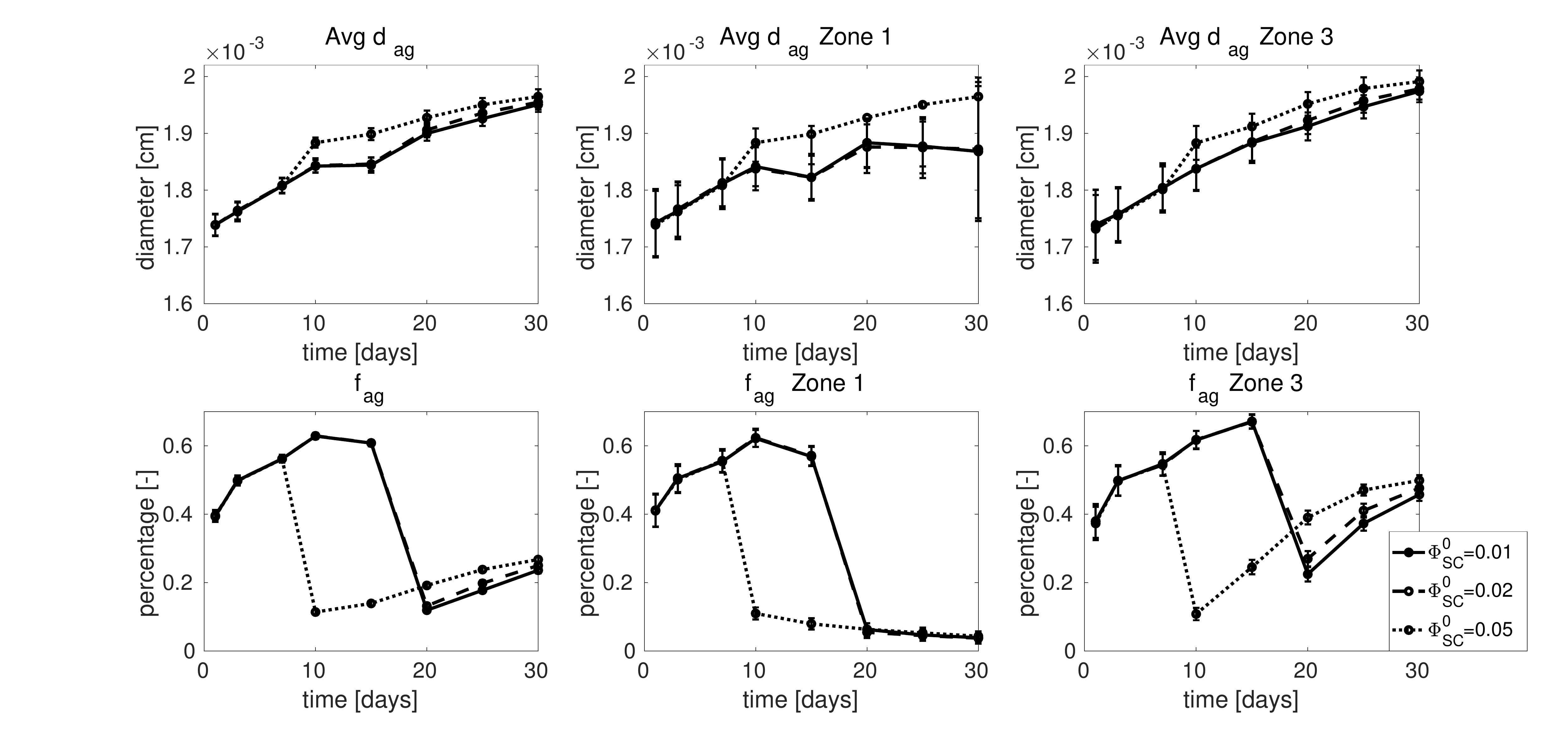}};
		\node (a) at (-4.5,2.3) {(a)};
		\node (b) at (-1.35,2.3) {(b)};
		\node (c) at (1.85,2.3) {(c)};
		\node (d) at (-4.5,-0.2) {(d)};
		\node (e) at (-1.35,-0.2) {(e)};
		\node (f) at (1.85,-0.2) {(f)};	
		\end{tikzpicture}
	\end{center}
	\caption{Model predictions of average diameter of cell aggregates, $d_{ag}$, on: (a) the entire domain $\Omega$, (b) Zone 1, and (c) Zone 3. The percentage of cells in aggregates, $f_{ag}$, on: (d) the entire domain $\Omega$, (e) Zone 1, and (f) Zone 3. All simulations reported here used the baseline parameters reported in Table~\ref{tab:par} and are for three different values of scaffold volume fractions: $\Phi^0_{SC}=0.01$ (solid line), 0.02 (dashed line), and 0.05 (dotted line). The average quantities and standard deviations are reported at $t=\{1,3,7,10,15,20,25,30\}$ days.}
	\label{fig:aggr}
\end{figure}

The presence of cell aggregates has been shown to influence several processes in the development of tissue engineered articular cartilage, thus we investigate  cell clustering in our simulations.
To further quantify the nature of cell aggregates obtained in the numerical simulations we compute the average cell aggregate diameter $d_{ag}$ and the average fraction of cells in aggregates, $f_{ag}$. Two cells are considered part of the same aggreagate if their center-to-center distance is less than $c_d$. The aggregate dimeter, $d_{ag}$, is computed as the maximum distance between the cell centers in the same aggregate plus $c_d$. The value is then averaged between all aggregates at a corresponding time step. The values depicted in Fig.~\ref{fig:aggr}(a)-(c) at each time point is the mean aggregate diameter $d_{ag}$ over 300 simulations and the standard deviation is shown at 8 time points. Isolated cells (aggregates of one cell) are not considered in the computation of the average $d_{ag}$. For different initial scaffold solid volume fraction $\Phi_{SC}^0$,  Fig.~\ref{fig:aggr} captures the average dynamics of cell agreggates on the entire domain in (a), Zone 1 in (b), and Zone 3 in (c). The model predicts an average value of $d_{ag}$ of 1.7-2 $\times 10^{-3}$cm, which is in agreement with the value of 20$\mu$m (2$\times 10^{-3}$cm) reported by \cite{Vunjak98}. As expected, cell aggregate diameter  $d_{ag}$ is growing in time in (a)-(c), consistent with increased cells in the construct. In addition, $\Phi_{SC}^0=0.05$ has aggregates that are larger than the other cases for days 10-30. Since Fig.~\ref{fig:aggr}(a)-(c) only captures the average diameter of cell clusters, we also wanted to investigate the fraction of cells in aggregates.

The fraction of cells in aggregates, $f_{ag}$, is computed as the ratio between the number of cell in aggregates and the total number of cells at a given time. The dynamics of the value of $f_{ag}$ are shown in Fig.~\ref{fig:aggr} for the entire domain in (d), Zone 1 in (e), and Zone 3 in (f). In (d)-(f), the fraction of cells in aggregates are characterized by an increase followed by a rapid decrease in all of the zones considered. There is a location specific behavior in Zone 1 and Zone 3. After the rapid decrease of $f_{ag}$ in Zone 1 shown in (e), it then remains almost constant  whereas it increases rapidly in Zone 3 in (f). The behavior in the whole domain $\Omega$, Fig.~\ref{fig:aggr}(d), seems to be intermediate between Zone 1 and Zone 3. The dynamics for both the aggregate size and fraction of cells in aggregates in Zone 1 and 3 are consistent with the data reported in Fig.~\ref{fig:barbaseline}. The uniform nutrient distribution applied at the beginning of the simulation promotes uniform cell growth in the whole domain, independent from location, and therefore more collisions and the increase in $f_{ag}$ reported in Fig.~\ref{fig:aggr}(d). When the cells reach a critical value of porosity, their movement is limited and they have to move smarter causing a decrease in collisions and thus a decrease in $f_{ag}$. Then, the lower nutrient concentration in Zone 1 will result in a lower number of cells and therefore a lower value of $f_{ag}$ as shown in (e), whereas the high nutrient concentration in Zone 3, shown in (f), will promote cell division and the increase of $f_{ag}$. We note that the temporal dynamics of the variations in $f_{ag}$ are faster for the simulations with a higher initial scaffold volume fraction ($\Phi_{SC}^0$). This is consistent with the dynamics depicted in Fig.~\ref{fig:baseline}, since a higher initial scaffold volume fraction corresponds to achieving the critical value of porosity at an earlier time point, triggering the qualitative change in behavior.

\begin{figure}[ht]
	\begin{center}
		\begin{tikzpicture}
		\node (fig) at (0,0) {	\includegraphics[width=0.95\textwidth]{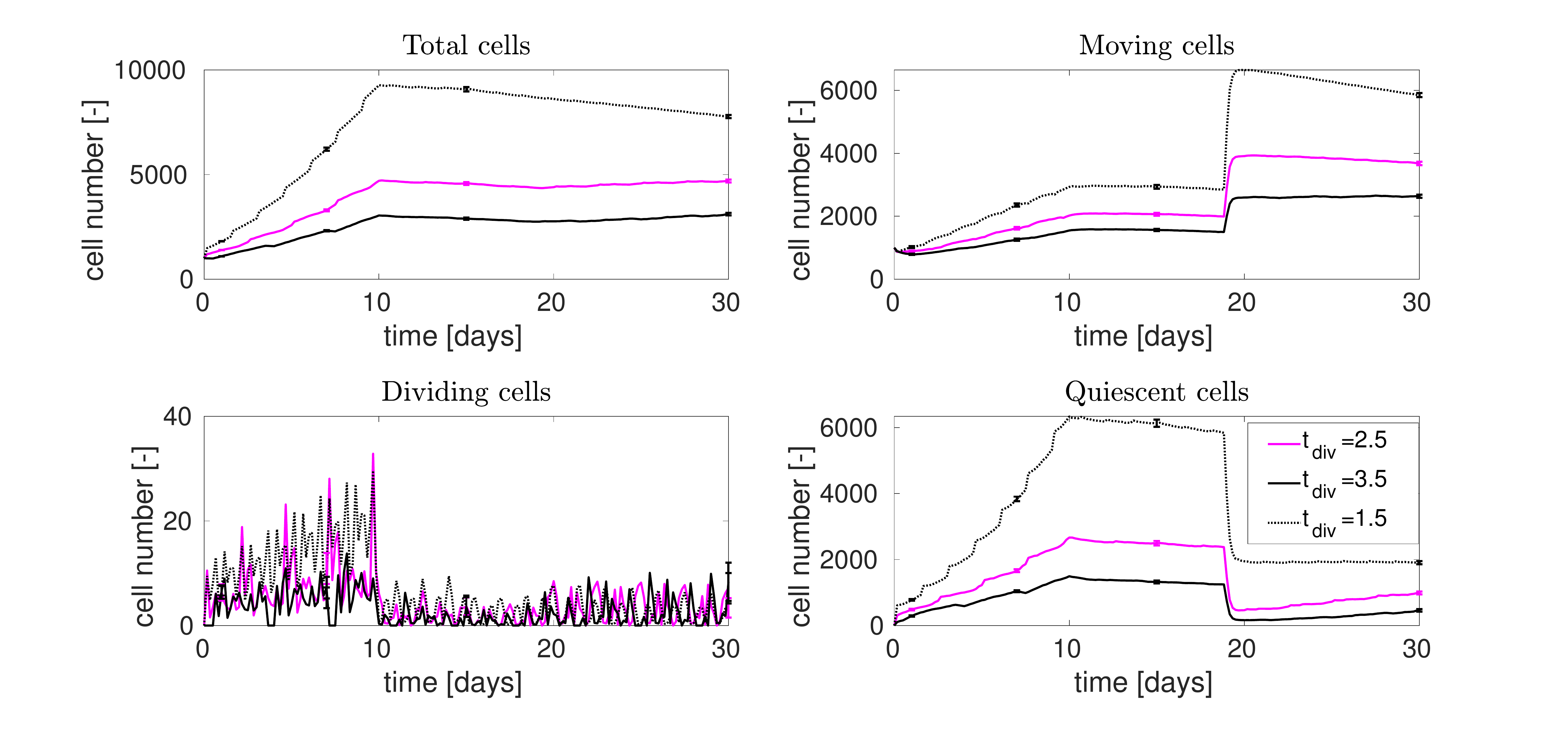}};
		\node (a) at (-4.5,2.4) {(a)};
		\node (b) at (0.5,2.4) {(b)};
		\node (c) at (-4.5,-0.1) {(c)};
		\node (d) at (0.5,-0.1) {(d)};
		\end{tikzpicture}
	\end{center}
	\caption{Total number of cells and cell counts for the different cell statuses as $t_{div}$, the time for a cell to mature and be able to divide, is varied. All other parameters are from Table \ref{tab:par} and initial scaffold volume fraction is $\Phi_{SC}^0=0.01$. The graph displays the total number of cells in (a) and cells in the following states: (b) moving, (c) dividing, and (d) quiescent. The standard deviation is shown at days 1, 7, 15, and 30. }
	\label{fig:maturity}
\end{figure}
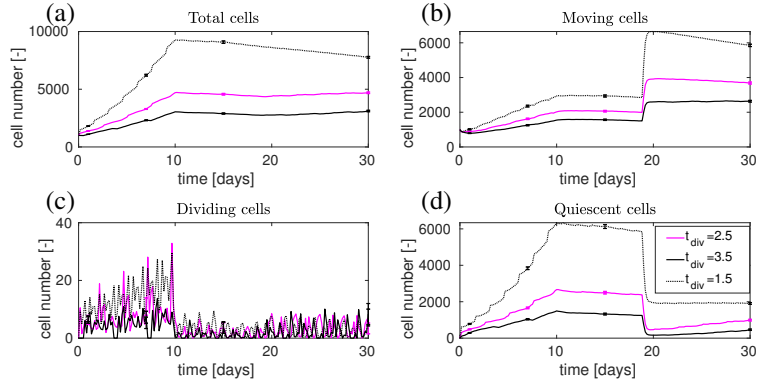

We note that cell characteristics such as movement, reproduction, and synthesis of extracellular matrix will depend heavily on the initial cell type as well as the type of scaffold that the cells are seeded in. Hence, parameters in this model can be tuned for different experimental setups and we can explore how variations in these parameter values affect the cellular distribution within the construct. In the literature, the average time for a chondrocyte to mature and be able to divide in different experimental conditions is estimated as 1.5-3 days and our baseline value was $t_{div}=2.5$ \citep{Bandeiras15,Chung10,Freed94}. We investigate increasing ($t_{div}=3.5$ days) and decreasing ($t_{div}=1.5$ days) the cell maturity time in Fig.~\ref{fig:maturity}. As expected, when $t_{div}$ is decreased and cells mature at a faster rate, the total cell number increases  (Fig.~\ref{fig:maturity}(a)) and we observe that more cells are undergoing division in the first 10 days (Fig.~\ref{fig:maturity}(c)). Correspondingly, there are more cells, which results in an increase in cell collisions and cells in the quiescent state as shown in (d). As cells can no longer undergo division since they do not meet the criteria in terms of porosity and nutrient concentration, cells divide less frequently after 10 days as shown in Fig.~\ref{fig:maturity}(c). We observe in Fig.~\ref{fig:maturity}(b) that cells move at a higher rate after day 20, since they are able to move throughout the domain and rearrange themselves in a way that collisions are greatly reduced as shown in (d). The proposed decrease in cell maturity age $t_{div}$ (faster maturity) resulted in a significant increase in total cells whereas the increase in $t_{div}$ (slower maturity) led to a  decrease in cell number relative to the baseline case. This is due to the fact that with the decrease in cells, there are fewer cells that can divide through the course of the simulation. We note that the spikes in the number of dividing cells shown in  Fig.~\ref{fig:maturity}(c) occurs at different time points, which is consistent with the different maturity time scales. We note that results in Fig.~\ref{fig:maturity} are shown for an initial scaffold solid volume fraction $\Phi_{SC}^0=0.01$ and that similar trends are observed for 0.02 and 0.05 (results not shown). The rapid change observed in Fig.~\ref{fig:maturity}(b) and (d) around day 18 happens at an earlier time point for higher initial scaffold volume fractions, consistent with the temporal dynamics in Fig.~\ref{fig:baseline}.

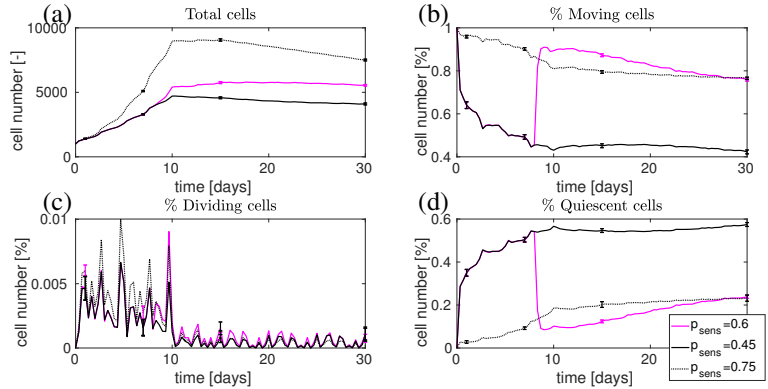
\begin{figure}[ht]
	\begin{center}
		\begin{tikzpicture}
		\node (fig) at (0,0) {	\includegraphics[width=0.95\textwidth]{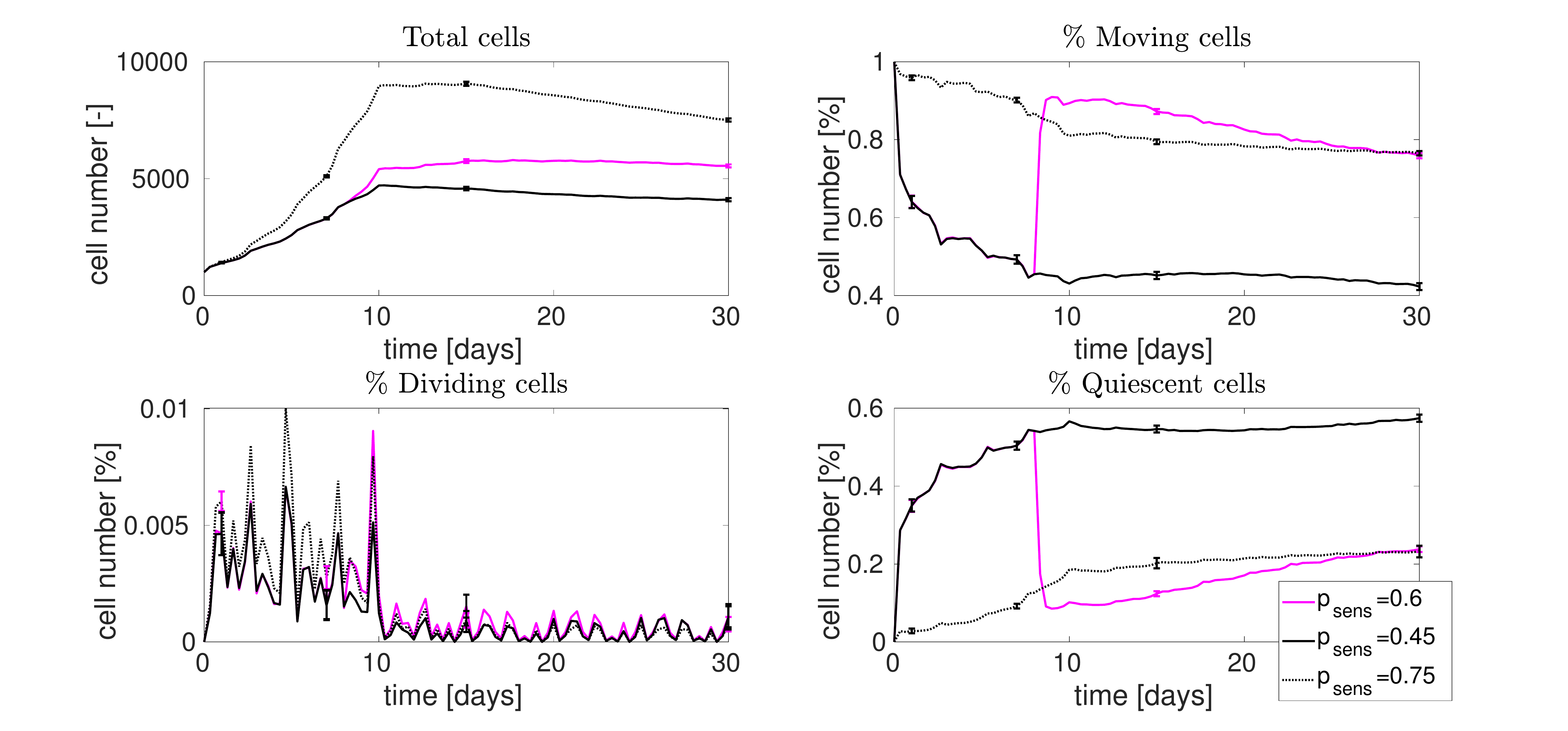}};
		\node (a) at (-4.5,2.4) {(a)};
		\node (b) at (0.5,2.4) {(b)};
		\node (c) at (-4.5,-0.1) {(c)};
		\node (d) at (0.5,-0.1) {(d)};
		\end{tikzpicture}
	\end{center}
	\caption{Cell counts and cell states as $p_{sens}$, the sensitivity to porosity, is varied for initial scaffold volume fraction $\Phi_{SC}^0=0.05$. All other parameters are from Table \ref{tab:par}. The graphs display the total number of cells in (a), the percentage of cells in the moving state in (b), the percentage of cells in the dividing state in (c), and the percentage of cells in the quiescent state in (d). The standard deviation is shown at days 1, 7, 15, and 30.}
	\label{fig:phi}
\end{figure}

In this model, we have biased cellular movement based on both the local nutrient concentration as well as the porosity. In the baseline parameters, we set the sensitivity to porosity to $p_{sens}=0.6$ such that a cell is not likely to move in the direction with porosity less than this value, and it won't divide in a region with porosity less than $p_{sens}=0.6$. Since we do not have an experimentally measured value for this parameter, we vary it higher and lower to understand how this changes the cellular dynamics in the construct. Results are shown in Fig.~\ref{fig:phi} for the case of initial scaffold volume fraction $\Phi_{SC}^0=0.05$ and a high and low sensitivity, corresponding to $p_{sens}=0.75$ and $p_{sens}=0.45$, respectively. Higher sensitivity actually leads to an increase in total cell count over the baseline case as shown in (a), which is counter intuitive at first glance since one would think that higher sensitivity might prevent movement and cell division. However, the higher sensitivity to porosity 
leads to increased cellular movement as shown in (b) and a decrease in quiescent cells as shown in (d), which results in decreased cell clustering and collisions. This then allows for more cells to divide, shown in (c), since they are more spread out. Lower sensitivity has total cell counts similar to the baseline for the first 8 days. After that time point, lower sensitivity to porosity has higher probability of movement leading to cell clustering and as a result, there is a much larger percentage of cells that collide and end up in the quiescent state. We note that there is a non symmetric behavior in that the same magnitude of increase and decrease in sensitivity to porosity gives changes in total cell number that are very different in magnitude. Similar trends are observed for $\Phi_{SC}^0=$0.02 and 0.01 (results not shown), and the rapid change observed in the magenta line ($p_{sens}=0.6$) in Fig.~\ref{fig:phi}(b) and (d) around day 8 happens at later time points for lower initial scaffold volume fractions, consistent with the temporal dynamics in Fig.~\ref{fig:baseline}.

\begin{figure}[ht]
	\begin{center}
		\begin{tikzpicture}
		\node (fig) at (0,0) {	\includegraphics[width=0.95\textwidth]{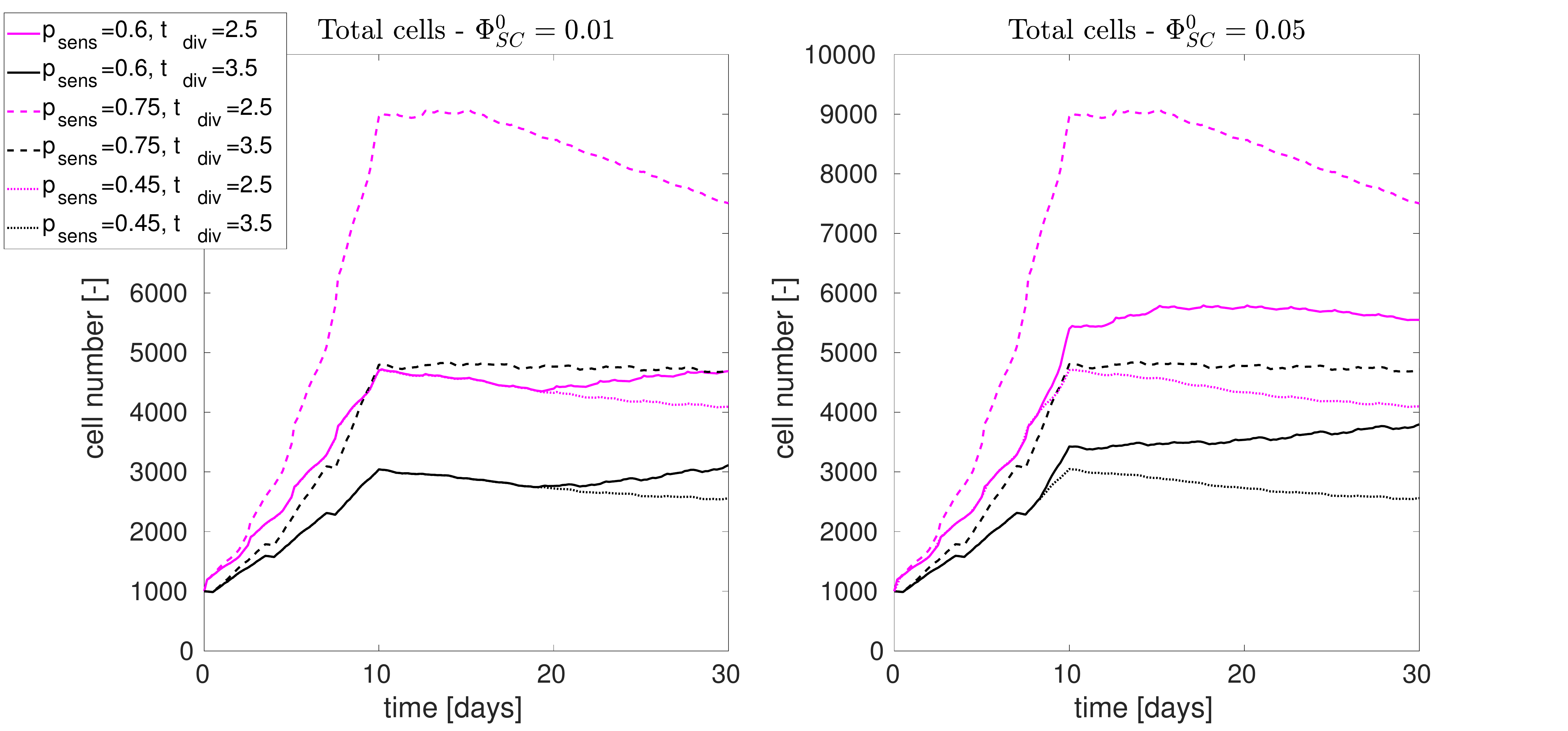}};
		\node (a) at (-4.5,2.7) {(a)};
		\node (b) at (0.5,2.7) {(b)};
		\end{tikzpicture}
	\end{center}
	\caption{Total cell number for initial scaffold volume fraction $\Phi_{SC}^0=0.01$ in (a) and $\Phi_{SC}^0=0.05$ in (b) when varying sensitivity to porosity $p_{sens}$ and cell maturity time $t_{div}$. All other parameters are specified in Table \ref{tab:par}. }
	\label{fig:phi-mat1p5p}
\end{figure}

To further understand the dynamics of cell counts in the bio-construct, we explore varying the cell maturity time $t_{div}$ and the sensitivity to porosity $p_{sens}$ in Fig.~\ref{fig:phi-mat1p5p}. In the case  of 1\% initial scaffold macromer concentration ($\Phi_{SC}^0=0.01$) in Fig.~\ref{fig:phi-mat1p5p}(a), when 
cell maturity time is increased, black lines, there is a consistent decrease in cell total for each porosity sensitivity, and the magnitude of the decrease is more pronounced for higher values of $p_{sens}$, dashed lines. The simulations with higher sensitivity are characterized by a higher growth rate which is greatly affected by an increase in $t_{div}$. However when there is a high sensitivity to porosity \textit{and} a longer cell maturity time ($t_{div}=3.5$ days), dashed black line, the system is still able to achieve a cell count close, and even higher at some time points, to simulations with lower $p_{sens}$ and shorter cell maturity time $t_{div}$, solid and dotted magenta lines. Cells with $p_{sens}$=0.75 are moving smarter than the cases with lower values of $p_{sens}$, and are therefore able to divide when they reach maturity, thus being able to fill the gap with the simulations with lower cell maturity time and obtain a similar total cell count. The effects of variations of $p_{sens}$ for the higher value of $t_{div}=$3.5 days is qualitatively similar to the variations obtained for $t_{div}=$2.5 days suggesting the importance of the parameter $p_{sens}$. 
In comparison, we can look at results for a higher initial scaffold volume fraction of 0.05 in Fig.~\ref{fig:phi-mat1p5p}(b). Similar to the 0.01 case, a higher sensitivity to porosity results in an increased cell count when the cellular maturity time is decreased. However, in the 0.05 case, we observe that cell counts are more dependent on the sensitivity to porosity in combination with cellular maturity time.

\begin{figure}[ht]
	\begin{center}
		\begin{tikzpicture}
		\node (fig) at (0,0) {	\includegraphics[width=0.95\textwidth]{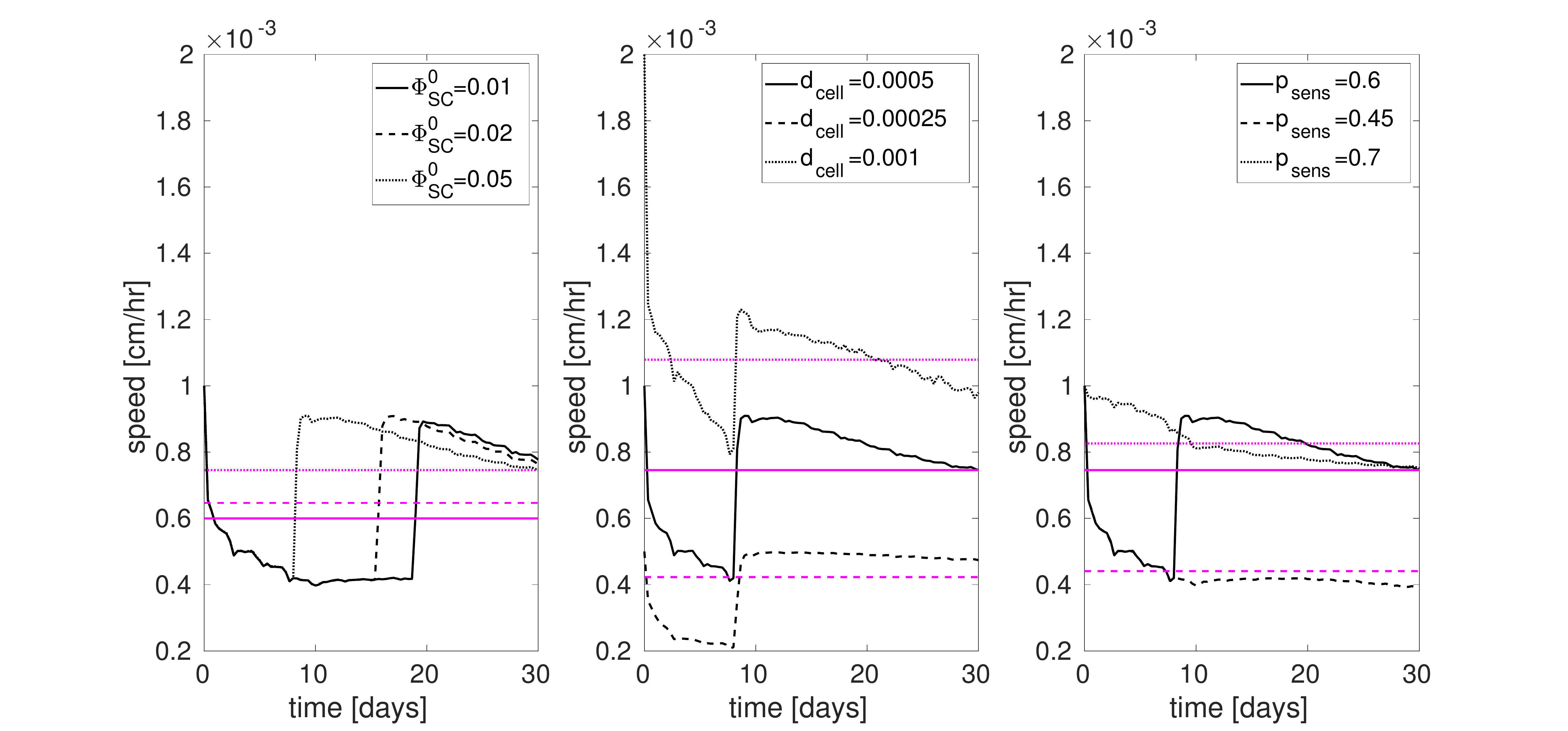}};
		\node (a) at (-4.5,2.5) {(a)};
		\node (b) at (-1.35,2.5) {(b)};
		\node (c) at (1.85,2.5) {(c)};
		\end{tikzpicture}
	\end{center}
	\caption{(a) Average cell speed for different values of initial scaffold volume fraction $\Phi_{SC}^0=0.01$ (solid black), $\Phi_{SC}^0=0.02$ (dashed black), and $\Phi_{SC}^0=0.05$ (dotted black). (b) Average cell speed for different values of $d_{cell}$ of 0.0005 cm (solid black), 0.00025 cm (dashed black), and 0.001 cm (dotted black) for $\Phi_{SC}^0=0.05$. (c) Average cell speed for different values of $p_{sens}=$ 0.6 (solid black), 0.45 (dashed black), and 0.75 (dotted black) for $\Phi_{SC}^0=0.05$. The magenta lines in (a)-(c) correspond to the average speed over 30 days for the corresponding black linestyles. The remaining parameters for each case are reported in Table~\ref{tab:par}.}
	\label{fig:speed}
\end{figure}
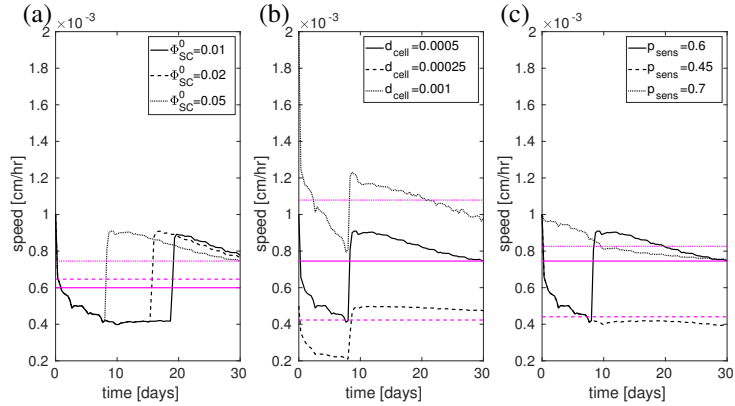

In our simulations, we fix the maximum distance that a cell may move at a given time iteration to $d_{cell}$.  If the rules allow for a cell to move, its velocity is then $v_c=d_{cell}/\triangle t=0.1\times 10^{-3}$cm per hour. However, not all cells move at each time step, so we can determine the emergent mean cell speeds, which we report in Fig.~\ref{fig:speed}(a). For all three initial scaffold solid volume fractions, there is a decrease in the emergent cell speed for the first 7 days. Around day 8, the $\Phi_{SC}^0=0.05$ case first starts to increase due to a decrease in quiescent cells where the 0.02 case follows around day 14 and the 0.01 case at day 18. The mean emergent cell speed over the course of the 30 day simulation is also shown (horizontal lines) for each of the different scaffold cases in Fig.~\ref{fig:speed}(a); we observe that the there is a monotonic trend of increasing mean cell speeds as initial scaffold solid volume fractions increases. \cite{Morales07} reviewed many experiments that studied chondrocyte movement under different experimental conditions. In general, there were two classes of cells, motile and nonmotile. The motile cells were reported as moving 5-10 $\mu$m per hour on average, with a maximum observation of 50 $\mu$m per hour. These cell speeds were recorded for movement on planar substrates, but fully three-dimensional movement in a scaffold could differ considerably. In addition, these experiments were carried out over shorter time periods whereas our simulation is for 30 days. Our emergent cell speeds are on the range of 6-7.5 $\mu$m per hour (0.6-0.75$\times 10^{-3}$cm per hour), so this is on the right order of magnitude. In Fig.~\ref{fig:speed}(b), we explore varying the maximum distance $d_{cell}$ that a cell can move for $\Phi_{SC}^0=0.05$. Here, a larger maximum distance a cell can move corresponds to a larger mean cell speed on the range of 0.1-1$\times 10^{-3}$cm per hour, which is similar to previous results of a cellular automaton model where porosity was not accounted for \citep{Chung10}. For the baseline distance  $d_{cell}$ we explore varying sensitivity to porosity $p_{sens}$ in Fig.~\ref{fig:speed}(c), again for the case of $\Phi_{SC}^0=0.05$. With a higher sensitivity to porosity ($p_{sens}=0.75$), cells move faster since there are more moving cells and less cells in the quiescent state. As aggregates of cells are forming, cell speed decreases and reaches the same value as the baseline porosity sensitivity, $p_{sens}=0.6$. Since the lower sensitivity to porosity ($p_{sens}=0.45$) allows cells to move less optimally, this causes an increased number of cells in the quiescent state, and results in an emergent cell speed that is approximately half that of the other cases. Again, we note that all of these cell speeds are in the experimentally observed range.

\section{Conclusion}
In this model, we have developed a framework to capture individual cellular interactions within a porous scaffold where cellular solid volume fraction is captured. Cell movement and proliferation is biased based on the continuously defined nutrient and porosity. Through a phenomonological model of ECM synthesis and scaffold degradation, we are able to account for these evolving solid volume fractions to update the local porosity. The model is parameterized and matched to experimental trends of total cell counts in cell-seeded scaffolds \citep{Freed94}. We are able to illustrate complex dynamics of cells in different states (moving, dying, proliferating, and quiescent) in both the total construct and in particular regions of the domain. Due to the fixed and high nutrient concentration at boundaries exposed to the nutrient bath, we observe aggregates of cells in these regions at later time points, which we characterize and match with reported aggregates in experiments \citep{Vunjak98}. In addition, we observe emergent cell speeds in the range reported for chondrocytes \citep{Morales07}. In this study, we have explored in detail, the emergent behavior of cells and total cells in the construct as the initial scaffold volume fraction, sensitivity to porosity, movement distance, and cell maturity time is varied. Model results show that cells with a higher sensitivity to porosity are able to move in a more optimal manner, decreasing cellular collisions and being able to proliferate more often, on average. As the cellular maturity time is decreased, this leads to a significant increase in cells in the construct. Each of these results can have a significant impact on understanding whether a growth factor or additional nutrients need to be supplied to the system, as well as determining the optimal scaffold volume fraction for a particular cell type. 

In \cite{Erickson09}, experiments were completed where mesenchymal stem cells were seeded in either agarose or  methacrylated hyaluronic acid (MeHA) scaffolds. The DNA per construct (measure of cell count in the construct) was measured as an average of 4 samples at day 1, 21, and 42. In this data, similar cell counts are observed at day 1 whereas the 1\% case had less DNA (and hence cells) at day 21 and 42 in comparison to the 2 and 5\% cases. Our simulations are consistent with this trend of the 1\% MeHA case ($\Phi_{SC}^0=0.01$) having less cells than the 5\% case ($\Phi_{SC}^0=0.05$) throughout the 30 day simulation. In comparison to the experiments of \cite{Erickson09}, we also observe a large amount of growth from day 1 to 21 but in contrast, we do not observe significant growth in days 21 to 30. We are not able to capture the larger difference between the 1 and 2\% cases, but this might be due to the use of different cell and scaffold type (stem cells and MeHA or agarose) in comparison to what we have parameterized our model to (chondrocytes and PGA) from the data of \cite{Freed94}.  With additional data of how different cell types move in different three-dimensional scaffolds, we will be able to parameterize this model to make predictions for different experiments.

In this model, there were several assumptions made that need to be considered when interpreting these results. We have accounted for ECM accumulation and scaffold degradation as being constant throughout the domain at each time point. However, from experiments of \cite{Erickson09}, there is evidence that matrix constituents are localized around cells, especially in the case of scaffolds with higher initial solid volume fraction. Additionally, since an outcome of interest is the mechanical properties of the tissue engineered construct, we would want to have a better representation of ECM accumulation, since this is been found to correlate with compressive moduli \citep{Bandeiras15,Cigan16}. The process of ECM formation, deposition, and transport has been previously analyzed by several continuum models \citep{Dimicco03,Nikolaev10,Obradovic00,Saha04,Sengers04} using a system of differential equations, but this presents challenges to couple to the cellular automaton approach where we are  modeling individual cells. In the future, it will be interesting to extend the work of \cite{Trewenack09}, which captured the distribution of ECM and scaffold around a single chondrocyte, to a discrete framework. 

The nutrient bath surrounding the construct was assumed to be well mixed, but we did not account for the effect of nutrient profusion on cell growth as in previous models \citep{Chung07,Hossain15,Nava13,Sacco11}. The nutrient concentration profile detailed in Appendix B is also independent of cell location. We account for increased cells in an ad hoc manner where exterior boundaries exposed to the fluid bath have a high nutrient concentration and the inner part of the construct has a decreasing concentration in time. Qualitatively, this provides enough input into the system to bias cell movement towards the edges of the construct. A more exact nutrient profile could be obtained by coupling a partial differential equation for the nutrient profile to the exact cell locations as in \cite{Chung10}. However, since we account for the evolving porosity, we would also need to include a diffusion constant that varies with porosities in extensions of this model. These model limitations will be the focus of future studies where we will also couple variable movement based on porosity, extend cell based rules to change behavior of individual cells in aggregates, and investigate cellular dynamics when nutrient channels or pores are included to deliver nutrients as in \cite{Chung10} and \cite{Cigan16}. \medskip

\noindent{\bf Acknowledgements:} We would like to acknowledge Mansoor Haider for useful discussions.\\
\textbf{Funding:} This study was not funded. \\
\textbf{Conflict of Interest:} The authors declare that they have no conflict of interest.

\section*{Appendix A: Numerical details}\label{app:numerics}

In the following section we provide additional details on some of the modeling choices made in this paper and for some of the parameters reported in Table~\ref{tab:par}.

The value of the cellular volume fraction at a node $\hat{\Phi}=0.3$ is determined based on a cell packing problem. Experiments have shown that chondrocytes are spherical in shape and may alter their shape based on the environment \citep{Hauselman92,Kino05}. We consider the cell as a sphere of diameter $c_d$ cm immersed in the thin three-dimensional domain with thickness $z=2c_d$ cm depicted in Fig~\ref{fig:sc}(b). The volume of the cell is 
$$V_c=\dfrac{4}{3}\pi \left(\dfrac{c_d}{2}\right)^3=\dfrac{c_d^3}{6}\pi \text{ cm}^3$$
and the cell is completely confined in a rectangular prism with dimensions ${c_d\times c_d\times 2c_d}$ and volume $V_r=2c_d^3$ cm$^3$. The ratio 
$$\dfrac{V_c}{V_r}=\dfrac{\pi}{12}\simeq 0.26$$
determines the volume fraction of the prism occupied by the sphere. Assuming the packing is not optimal, because of deformation limits of the cells, the value of $\hat{\Phi}$ is rounded up to $0.3$. The value is applied to every grid point covered by the cell surface represented by the gray shaded area in Fig.~\ref{fig:cell-nbhd}. Consequently, the value of the sensitivity to porosity is chosen as $p_{sens}=2\hat{\Phi}=0.6$ so that a cell will not move or divide in a region that could not accommodate two full cells. This choice is made to provide some extra room for cells in case of a simultaneous movement of two nearby cells into the same region or cell division happening on the same target neighborhood. This is also done to have a  maximum number of cells that could fit on the whole domain, which is comparible with biological experiments.

The outer diameter of the annulus $a_{d,mov}$ represents how far the chondrocyte cell is able to detect gradients in nutrient concentration and porosity, and it is chosen in this work to be six times the baseline cell movement ($6d_{cell}=3c_d$) in one iteration. The outer diameter of the annulus $a_{d,div}$ is chosen equal to $3.5c_d$ since a gap of one grid point between the mother cell and the new cell is needed to avoid overlaps during cell division.

\section*{Appendix B: Nutrient profiles}\label{app:nutrient}
Here we detail the explicit formulation for the nutrient profiles used in the numerical simulations. The profiles are chosen to qualitatively agree with the ones reported in~\cite{Chung10} and~\cite{Bandeiras15}. In this work the nutrient profiles are updated at day 0, 7, 10, and 15 on the simulation domain ${\Omega=[0,0.5]\times[0,0.1]}$, with the points (0,0) and (0.5,0.1) corresponding to the top left and bottom right corners of the domain, respectively. At day 0 the scaffold is well perfused, due to the high porosity, and the nutrient concentration is assumed uniform for several days where
\begin{equation*}
c_0(x,y,\hat{t}) = 1, 
\end{equation*}
for $(x,y)\in\Omega$ and for $\hat{t}\in[0,7)$ where $\hat{t}$ corresponds to days. At day 7, the concentration in the inner portion of the domain starts to decrease due to cellular consumption of nutrients and the profile is updated as follows
\begin{equation*}
c_7(x,y,\hat{t}) = \left\{ 
\begin{array}{ll}
0.75, \qquad & \left(\frac{x-0.3}{0.06}\right)^2 + \left(\frac{y}{0.012}\right)^2\leq 1, \\
\\
0.83, & \left(\frac{x-0.3}{0.06}\right)^2 + \left(\frac{y}{0.012}\right)^2>1, \text{ and } \left(\frac{x}{0.47}\right)^2 + \left(\frac{y}{0.064}\right)^2<1,\\
\\
1, & \left(\frac{x}{0.47}\right)^2 + \left(\frac{y}{0.064}\right)^2\geq 1.
\end{array}
\right.
\end{equation*}
for $(x,y)\in\Omega$ and $\hat{t}\in[7,10)$. At day 10, the concentration in the inner domain keeps decreasing due to increased cellular consumption of nutrients and decreases in porosity on the outer part of the construct (the bottom and right hand sides). The nutrient concentration profile is updated using the following equation
\begin{equation*}
c_{10}(x,y,\hat{t}) = \left\{ 
\begin{array}{ll}
0.1, & \left(\frac{x-0.29}{0.14}\right)^2 + \left(\frac{y}{0.065}\right)^2\leq 1, \\
\\
0.2, & \left(\frac{x-0.29}{0.14}\right)^2 + \left(\frac{y}{0.065}\right)^2>1, \text{ and } \left(\frac{x-0.1}{0.35}\right)^4 + \left(\frac{y+0.02}{0.095}\right)^4\geq 1,\\
\\
\max\left\{ \min\left[ g_1(x,y) ,1\right],0.2\right\}, \quad & \left(\frac{x-0.1}{0.35}\right)^4 + \left(\frac{y+0.02}{0.095}\right)^4>1,
\end{array}
\right.
\end{equation*}
for $(x,y)\in\Omega$ and $\hat{t}\in[10,15)$ days. The rescaled gradient function $g_1$ is obtained as follows
$$
g_1(x,y)= 0.8 \, \dfrac{\left(\frac{x-0.15}{0.35}\right)^4 + \left(\frac{y}{0.095}\right)^4 - 0.54}{1-0.54} + 0.2.
$$
The decreased nutrient concentration in the upper left portion of the domain $\Omega$ is more pronounced at day 15 due to the dense layer of cells near the right and bottom boundary reducing nutrient diffusion to the center of the construct. Hence, the region with low nutrient in the inner portion of the domain increases and the region with a nutrient gradient (function $g_2$) gets thinner. The nutrient concentration is updated using the following equation
\begin{equation*}
c_{15}(x,y,\hat{t}) = \left\{ 
\begin{array}{ll}
0.1, & \left(\frac{x+0.05}{0.0.5}\right)^5 + \left(\frac{y+0.01}{0.1}\right)^5\leq 1, \\
\\
\max\left\{ \min\left[ g_2(x,y) ,1\right],0.1\right\}, \quad & \left(\frac{x+0.05}{0.0.5}\right)^5 + \left(\frac{y+0.01}{0.1}\right)^5   >1,
\end{array}
\right.
\end{equation*}
for $(x,y)\in\Omega$ and $\hat{t}\in[15,30)$ where the rescaled gradient function $g_2$ is obtained as follows
$$
g_2(x,y)= 0.9 \, \dfrac{\left(\frac{x}{0.5}\right)^5 + \left(\frac{y}{0.1}\right)^{5} - 0.59}{1-0.59} + 0.1.
$$

\begin{figure}[b!]
	\begin{center}
		\includegraphics[width=0.9\textwidth]{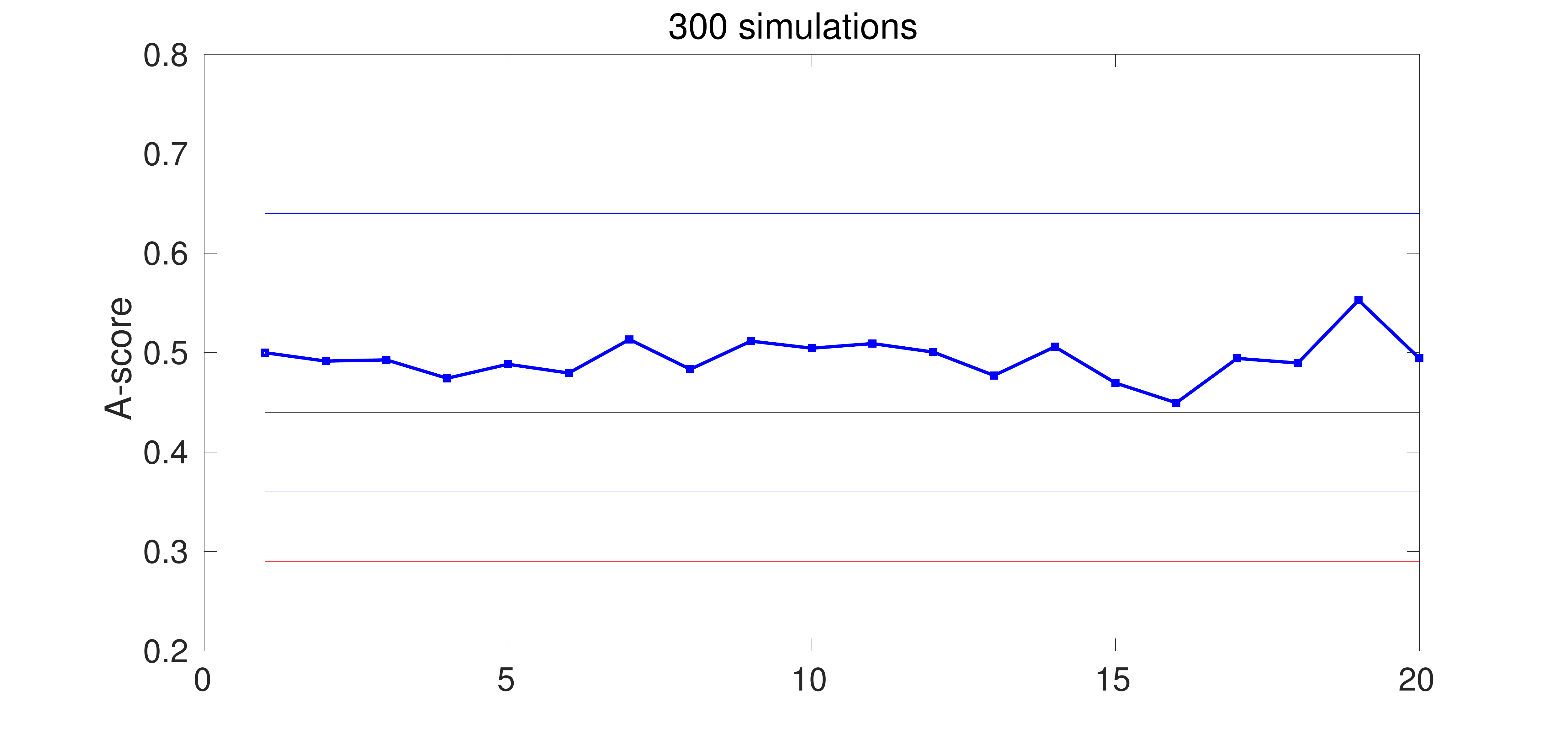}
	\end{center}
	\caption{A-score $A_{1,j}$ for comparison of set 1 with set $j=1,\ldots,20$ on the variable {\it final number of cells} in simulations with $\Phi_{SC}^0=0.01$ and parameters from Table~\ref{tab:par}.}
	\label{fig:ascore}
\end{figure}

\section*{Appendix C: Aleatory uncertainty analysis}\label{app:aleatory}

The model contains several stochastic elements that introduce aleatory uncertainty in each single simulation. It is therefore necessary to perform an uncertainty quantification procedure to determine the minimum number of simulations to account for the full variation of the system~\citep{Alden13,Cosgrove15,Read12}. First, 20 sets of $k$ simulations each with the same parameters are gathered. Then, sets 2-20 are compared with set 1 using the A-Test developed by~\cite{Vargha00}. The A-Test returns an A-score for each comparison which represents the probability that a randomly selected sample from the first population is larger than a random sample from the second population. The A-score for the comparison between set $a$ and $b$, $A_{a,b}$ is computed following the approximation proposed in~\cite{Vargha00} as:
\begin{equation*}
A_{a,b}=\dfrac{\#(X_i>Y_j)}{k_a k_b} \, + \, 0.5\,  \dfrac{\#(X_i=Y_j)}{k_a k_b}, \qquad i=1,\ldots,k_a; \, j=1,\ldots,k_b
\label{eq:ascore}
\end{equation*}
where $\#(X_i>Y_j)$ counts the number of times the event $(X_i>Y_j)$ happens for all the $X_i$ in set $a$ and $Y_j$ in set $b$ and similarly $\#(X_i=Y_j)$ counts the number of times the event $(X_i=Y_j)$ happens, and $k_a$ and $k_b$ are the dimension of set $a$ and $b$, respectively. The A-score is used to determine statistical significance as follows: a score below 0.29  or above 0.71 indicates a large effect of sample size $k$ on the model results; a score of 0.36 or 0.64 indicates a medium effect of $k$ on model results; a score within 0.44 and 0.56 indicates a small effect of $k$ on model results, and a score of 0.5 indicates no effect of $k$ on model results. Fig.~\ref{fig:ascore} shows the results obtained for sample size $k=300$ considering the final number of cells as the variable of interest, for simulations with the baseline parameters in Table \ref{tab:par} with $\Phi_{SC}^0=0.01$. 20 groups with $k=$300 distinct simulations are obtained and then the A-score $A_{1,j}$ is computed for $j=1,\ldots,20$ and reported in Fig.~\ref{fig:ascore}. The results obtained, with all of the A-scores within the $[0.44,0.56]$ interval, show that a value of 300 simulations is sufficient to mitigate the uncertainty introduced by the stochasticity of the model. The correct application of the A-Test is validated by the first point in Fig.~\ref{fig:ascore}, where the comparison of sample 1 with itself results in an A-score $A_{1,1}=0.5$ as expected from the literature.

\end{document}